\documentclass[prd,aps, preprint, nofootinbib,superscriptaddress]{revtex4-1}
\usepackage{amsmath}
\usepackage{epsfig}
\usepackage{subfigure}

\begin{document}

\hfill{NPAC 12-09}

\title{Supersymmetric Electroweak Baryogenesis\\ Via Resonant Sfermion Sources}

\author{Jonathan Kozaczuk}
\email{jkozaczu@ucsc.edu}\affiliation{Department of Physics, University of California, 1156 High St., Santa Cruz, CA 95064, USA}

\author{Stefano Profumo}
\email{profumo@scipp.ucsc.edu}\affiliation{Department of Physics, University of California, 1156 High St., Santa Cruz, CA 95064, USA}\affiliation{Santa Cruz Institute for Particle Physics, Santa Cruz, CA 95064, USA} 

\author{Michael J. Ramsey-Musolf}
\email{mjrm@physics.wisc.edu} \affiliation{University of Wisconsin-Madison, Department of Physics
1150 University Avenue, Madison, WI 53706, USA}
\affiliation{Kellogg Radiation Laboratory, California Institute of Technology, Pasadena, CA 91125 USA}

\author{Carroll L. Wainwright}
\email{cwainwri@ucsc.edu} \affiliation{Department of Physics, University of California, 1156 High St., Santa Cruz, CA 95064, USA}

\date{\today}

\begin{abstract}
\noindent We calculate the baryon asymmetry produced at the electroweak phase transition by quasi-degenerate third generation sfermions in the minimal supersymmetric extension of the Standard Model. We evaluate constraints from Higgs searches, from collider searches for supersymmetric particles, and from null searches for the permanent electric dipole moment (EDM) of the electron, of the neutron and of atoms. We find that resonant sfermion sources can in principle provide a large enough baryon asymmetry in various corners of the sfermion parameter space, and we focus, in particular, on the case of large $\tan\beta$, where third-generation down-type (s)fermions become relevant. We show that in the case of stop and sbottom sources, the viable parameter space is ruled out by constraints from the non-observation of the Mercury EDM. We introduce a new class of CP violating sources, quasi-degenerate staus, that escapes current EDM constraints while providing large enough net chiral currents to achieve successful ``slepton-mediated'' electroweak baryogenesis. 
\end{abstract}

\maketitle

\section{Introduction}

The origin of the matter-antimatter asymmetry observed in the universe ({\em baryogenesis}) is one of the greatest puzzles of modern cosmology. No ``standard model'' of baryogenesis exists, nor does it seem possible to accommodate baryogenesis within the Standard Model of particle physics \cite{Shaposhnikov:1986jp}, although numerous hypothetical frameworks have been investigated within various extensions to the Standard Model (see e.g. Refs.~\cite{Dine:2003ax, Morrissey:2012db} for reviews).

Of all the known options for a dynamical generation of the matter-antimatter asymmetry, one that may soon be conclusively tested is the framework of electroweak baryogenesis (EWB). This scenario generically entails the existence of particles that have electroweak quantum numbers and that are light enough to be in thermal equilibrium at the electroweak phase transition, when the universe had a temperature of around 100 GeV. A host of phenomenological consequences then follows:

(i) light, electroweakly-interacting particles should be produced at the Large Hadron Collider (LHC), making EWB testable with colliders (see e.g. Ref. \cite{Carena:2002ss, Carena:2008vj});

(ii) large enough CP violation and light enough new electroweakly-interacting particles generically produce sizable electric dipole moments (EDMs), via one- or two-loop effects, to the level probed, or close to being probed, by experiments (for recent studies see e.g. Ref.~\cite{EWB_and_EDMs, Li:2008kz});

(iii) if the lightest electroweakly-interacting particle is stable, it could be the dark matter; recent studies \cite{Balazs:2004ae, EWB_and_DM, Kozaczuk:2011vr} indicate that if this is the case, the resulting dark matter particle should be detectable with the current generation of dark matter search experiments;

(iv) bubble collisions at the necessarily strongly first-order EW phase transition produce gravity waves with characteristic frequency $f_{\rm EW}\gtrsim T_{\rm EW}/M_P$, which, today, have a frequency $f^{\rm today}_{\rm EW}\gtrsim 0.01$ mHz, in the range where the sensitivity of planned space-based interferometers peaks (see e.g. Ref.~\cite{Grojean:2006bp, Huber:2008hg, Caprini:2009yp}).

Broadly, the above phenomenological probes make EWB a testable scenario, even though quantitative predictions necessarily depend on the specifics of the model under consideration. For several reasons, independent of baryogenesis, a supersymmetric extension to the Standard Model has long been advocated as a particularly compelling one, with the minimal supersymmetric extension (MSSM) its simplest paradigm.

The MSSM contains enough degrees of freedom to overcome the two key difficulties encountered in embedding EWB within the Standard Model: (i) producing a strongly first order phase transition to prevent baryon number wash-out, and (ii) providing large enough sources of CP violation to quantitatively explain the observed baryon asymmetry. The first difficulty is overcome either by advocating a light scalar top (hereafter ``stop'') (there are numerous studies in this vein, from Ref.~\cite{Giudice:1992hh} to Ref.~\cite{Carena:2008vj}), by invoking a non-minimal Higgs sector comprising additional singlet fields (see e.g. Ref.~\cite{Menon:2004wv, Huber:2006wf}  and references therein), or by advocating higher-dimensional operators (e.g. Ref. ~\cite{Blum:2008ym} and references therein).  With regards to (ii), additional CP violating sources are also easily found in the MSSM parameter space.

Currently, the electroweak scale is under close scrutiny at the LHC, and EWB is either on the brink of discovery or of being convincingly ruled out, at least in its simplest incarnations. To address the question of whether or not EWB might be nature's mechanism of choice to produce the baryon asymmetry of the Universe (BAU), it is imperative to investigate in great detail all possible open corners of the MSSM parameter space that could be related to this scenario. Recent important progress has been made in evaluating the impact of the LHC program on point (i) above, with strong implications for the light-stop scenario \cite{Curtin:2012aa} but also for frameworks with non-minimal Higgs sectors \cite{Cohen:2012zt}. In the present study, we concern ourselves, instead, with issue (ii) above: which MSSM sector could have provided the CP violating sources necessary to feed the production of a baryon asymmetry at the electroweak phase transition? We leave aside here the (equally interesting, but separate) question of the mechanism responsible for producing a strongly enough first-order electroweak phase transition. The present study is intended to provide a qualitative and quantitative step forward in closing in on the allowed parameter space for supersymmetric EWB models.

Supersymmetric EWB has been extensively studied in many different contexts in the recent and not-so-recent past (see e.g. Refs.~\cite{Huet:1995sh, Lee:2004we, Chung:2008aya, Lepton_Mediated, Supergauge, Including_Yukawa, Carena:2002ss, Carena:2008vj, Konstandin:2004gy, Konstandin:2003dx, More_Relaxed, Balazs:2004ae, Huber:2001xf, CPV_and_EWB, Cline:2000kb, Chung:2009qs, Blum:2010by}).  
Most previous studies of EWB in the MSSM have insisted and focused on the  CP violation relevant to produce the baryon asymmetry as occurring in the higgsino-gaugino sector. This scenario generically forces a close degeneracy between one (or both) of the electroweak gaugino soft supersymmetry breaking masses $M_1$, $M_2$, and the supersymmetric $\mu$ parameter, perhaps raising concerns about naturalness since $M_{1,2}$ and $\mu$ have entirely disconnected origins. Additionally, electroweak two-loop contributions to EDMs (see Ref.~\cite{Li:2008kz} for the complete set of relevant two-loop contributions) provide stringent constraints on the viable parameter space.

Perhaps more importantly, there does not need to be any CP violation in the gaugino-higgsino ``fermionic'' supersymmetric sector for EWB to be successful. As noted long ago \cite{Carena:2002ss}, CP violation in the stop sector is potentially an important source for EWB in the MSSM (and beyond \cite{Blum:2010by}). Stop sources have not received much attention in previous studies primarily because of the superficial tension that typically exists between producing the correct BAU and having a  phenomenology consistent with observation.  This point can be illustrated in short as follows: the phase transition in the MSSM is typically made strongly first order by a light right-handed stop (typically in the sub-100 GeV range -- for an updated analysis see Ref. \cite{Carena:2008vj}); since the stop CP-violating sources are resonant for $m_{\tilde{i}_R}\sim m_{\tilde{i}_L}$, the parameter space most likely to produce the observed BAU has light, nearly degenerate stops.  However, light stops are tightly constrained by the Higgs mass, electroweak precision observables, and electric dipole moments (EDMs), along with null results from collider searches, thus making it difficult to have successful stop-sourced EWB along with a strongly first-order phase transition.  

If the ratio $\tan\beta$ of the two Higgs vacuum expectation values in the MSSM is sufficiently large, the Yukawa couplings of the down-type third generation fermions are generically of the same order as the top Yukawa coupling, therefore non-negligibly interacting with the Higgs fields. If this is the case, one would expect new contributions to the CP violating sources responsible for the generation of a net charge density of left-handed fermions, in turn fueling baryon number density production via weak sphalerons at the EW phase transition \cite{Chung:2008aya}. 
Unfortunately, down-type fermion superfields do not contribute to the strength of the EW phase transition, since the cubic term they produce in the effective potential is counter-balanced by a $1/\tan\beta$ factor, making down-type contributions smaller than the top contributions by factors of $m_{b,\tau}/m_t$. Nevertheless, CP sources relevant for EWB from scalar bottom and $\tau$ are generically significant in the large $\tan\beta$ regime. In this work, for the first time, we quantitatively study these sources, and we assess the impact of EDMs if CP violation is indeed sourced in the down-type (s)fermion sector.

This manuscript is structured as follows: we first elucidate, in Sec.~\ref{sec:asymmetry} the physical modeling of the electroweak phase transition relevant for the calculation of the baryon asymmetry of the universe (BAU), including details of the approximations we make and of the associated expected systematic effects. We then focus on stop sources in Sec.~\ref{sec:stops}, and study several slices of the relevant parameter space. We consider constraints from Higgs searches, EW precision observables, direct searches for supersymmetric particles with the LHC, and we devote special attention to constraints from EDM searches. Sec.~\ref{sec:largetanbeta} then studies the large $\tan\beta$ regime and the new class of CP violating sources introduced in this study: quasi-degenerate sbottoms and staus. As for stops sources, we investigate all relevant constraints, with special emphasis on EDM searches.

The main result of the present study is that the current EDM search limits -- particularly the one obtained for the Mercury atom \cite{Griffith:2009zz} -- eliminate the possibility that CP violating sources stemming from light and/or quasi-degenerate stops or sbottoms could be the main triggers for successful EWB. On the other hand, a new class of CP violating sources  associated with third generation sleptons is subject to considerably weaker EDM constraints. Consequently, these sources can successfully generate the production of the net left-handed chiral charge needed to produce the observed baryon asymmetry at the EW phase transition. This \lq\lq stau-mediated EWB" is possible, however, only in relatively narrow strips of parameter space where the two stau mass eigenstates are almost degenerate. 

In view of the fast pace with which the LHC is exploring the electroweak scale, and especially the supersymmetric sfermion sector, evaluating the potential relevance of sfermions to produce the observed baryon asymmetry appears to us as a timely topic. In addition, the program of searches for EDMs at the ``intensity frontier'' is also here demonstrated to be highly synergistic to the collider ``energy frontier''; searches for the EDM of multiple particle and atomic species is also crucial to testing the EW route to baryogenesis. Finally, while it is too early to draw strong conclusions, LHC searches for the Higgs might indirectly point to a scenario with light staus (see e.g. \cite{carlos}), potentially making the new source class we discovered all the more appealing.

\section{The Baryon Asymmetry in Supersymmetric Electroweak Baryogenesis}\label{sec:asymmetry}

In the framework of supersymmetric EWB, the baryon asymmetry of the universe is produced by $SU(2)$ sphalerons acting on a charge density of left-handed fermions, generated by CP-violating sources $S_i^{CPV}$ associated with the electroweak phase transition.  As bubbles of broken electroweak symmetry nucleate and expand, the CP-violating phases between the supersymmetric particles and the background Higgs fields lead to the production of net charge densities when (s)particles scatter off of the EWPT bubble wall.   In the present study we concern ourselves with scalar sources associated with third-generation (s)quarks and (s)leptons, as their Yukawa couplings are much larger than those of their first- and second-generation counterparts. In addition, we focus on the large $\tan\beta$ regime, where third-generation down-type Yukawa couplings become comparable to the top Yukawa coupling, and therefore relevant in scattering off of Higgs fields. The relevant part of the MSSM Lagrangian describing the associated CP-violating interactions in the gauge eigenstate basis reads: \begin{align}  \label{eq:lagrangian} \mathcal{L} \supset \hspace{3mm} &y_t \tilde{t}_L \tilde{t}^*_R (A_t H^0_u - \mu^*H^{0*}_d) +  y_b\tilde{b}_L \tilde{b}^*_R (A_b H^0_d - \mu^*H^{0*}_u) \\ \nonumber &+  y_{\tau}\tilde{\tau}_L \tilde{\tau}^*_R (A_{\tau} H^0_d - \mu^*H^{0*}_u)-bH^0_uH^0_d+ h.c., \end{align} where CP-violating phases can arise between the various triscalar couplings $A_f$, $\mu$, and the Higgs soft mass parameter $b$. We henceforth denote this phase for species $f$ as $\phi_f\equiv\mathrm{Arg}(\mu A_f b^\ast)$. Without loss of generality, we will assume $b$ to be real so that $\phi_f=\mathrm{Arg}(\mu A_f)$ in what follows.

In addition to the CP-violating sources, there are several CP-conserving processes arising from Eq.~(\ref{eq:lagrangian}) that affect particle number $n_i$ for the relevant species in the MSSM.  There are relaxation terms associated with chirality-changing particle scattering off of the Higgs vevs, with corresponding thermally-averaged rates $\Gamma^M_i$.   There are triscalar and supersymmetric Yukawa interactions given by Eq.~(\ref{eq:lagrangian}) without replacing $H^0_{u,d}$ by their vevs; as discussed below, the assumption of supergauge equilibrium allows us to combine the rates for both types of processes which we write as $\Gamma_{yi}$.  For the squarks, there are also $SU(3)$ sphalerons, with rate $\Gamma_{ss}$, that produce 1st- and 2nd-generation squarks from a 3rd-generation density and vice-versa.  Finally, weak sphalerons ultimately convert the left-handed particle density $n_L$ to a net baryon asymmetry with rate $\Gamma_{ws}$.  A complete set of expressions for these various sources can be found in Refs.~\cite{Lepton_Mediated, Chung:2009qs}, to which we refer the Reader for additional details of the calculation.  

We follow Refs.~\cite{Lee:2004we, Chung:2008aya, Lepton_Mediated, Supergauge, Including_Yukawa, Chung:2009qs} and work in the Higgs vev-insertion approximation, in which it is assumed that the sources in the bubble wall are strongest near the unbroken phase and where one uses a basis of $SU(2)$ gauge eigenstates, expanding about flavor-diagonal states in the bubble wall.  This approximation tends to overestimate the resulting baryon asymmetry and clearly breaks down farther inside the wall where flavor mixing cannot be neglected. However the vev-insertion approximation is expected to characterize the production of the BAU to order unity accuracy \cite{Chung:2009qs}.  Recent studies have worked out the flavor oscillations in the bubble wall beyond the vev-insertion approximation for a toy model \cite{Cirigliano:2011di}, and found qualitatively similar results to those obtained in the vev-insertion approximation, including a resonance in the various sources.  Although a treatment beyond the vev-insertion approximation is desirable for an accurate assessment of EWB in the MSSM, since we will be looking at the baryon asymmetry across a wide range of parameter space with other inherent uncertainties, we content ourselves with the vev-insertion approximation, deferring a more detailed analysis including flavor-mixing effects to future study.  Our results can thus be interpreted as a ``best case scenario" for EWB with scalar sources in the MSSM, albeit we also show results that would correspond to a factor 10 smaller net BAU, to guide the reader to a more conservative interpretation.

Proceeding within the outlined framework for computing the baryon-to-entropy ratio $Y_B$, the weak sphaleron rate $\Gamma_{ws}$ is typically much slower than the rates for the creation and diffusion of the left-handed charge density $n_L$ ahead of the EWPT bubble wall.  This allows us to consider separately the diffusion equations for the various (s)particle densities and the creation of the baryon density $n_B$, which is given, in terms of $\Gamma_{ws}$, $n_L$, and the bubble wall velocity $v_w$ as \cite{Huet:1995sh}: \begin{equation} \label{eq:BAU} n_B= \frac{-3 \Gamma_{ws}}{v_w} \int_{-\infty}^0 dz \hspace{1.5mm} n_L(z)e^{\frac{15\Gamma_{ws}}{4 v_w} z}, \end{equation}    where $z$ is the distance from the bubble wall in the wall rest frame (neglecting the wall curvature) and where the unbroken EW phase corresponds to $z<0$.  The left handed charge density $n_L$ is given by the sum of the charge densities of the various left-handed quarks and leptons $n_L=\sum(q_i+l_i)$ where the sum runs over all colors and generations and $q_i$, $l_i$ denote the difference of particle and antiparticle densities for each species.  The charge densities entering into the expression for $n_L$ are obtained from a set of coupled quantum Boltzmann equations (described below) which, once solved, allow one to compute $n_B$ via Eq.~(\ref{eq:BAU}).

Detailed derivations of the quantum Boltzmann equations (QBEs) governing the generation of the BAU have been discussed at length in the existing literature (see e.g. Ref.~\cite{Chung:2009qs} for a full treatment) so we do not reproduce them here; in what follows we use the simplified form of the QBEs discussed in Ref.~\cite{Lepton_Mediated}, with some modifications.  For each particle species we can define a corresponding chemical potential $\mu_i$, which is the fundamental quantity entering into the Boltzmann equations, related to its corresponding charge density by \begin{equation} n_i=\frac{T^2}{6}k_i \mu_i +\mathcal{O}\left(\frac{\mu_i}{T} \right)^3, \end{equation} where we have expanded in $\mu/T$ and the statistical weight for the density $k_i$ is given by \begin{equation} k_i=g_i \frac{6}{\pi^2}\int_{m_i/T}^{\infty}dx \frac{x e^x }{(e^x\pm1)^2}\sqrt{x^2-m_i^2/T^2}. \end{equation}  Additionally, as we will see in the following sections, for the cases we consider, the so-called supergauge rates, which drive chemical equilibrium between particles and their superpartners $\mu_i \leftrightarrow \mu_{\tilde{i}}$, are typically faster than the corresponding diffusion timescale $\tau_{\rm diff}$, defined in terms of the various diffusion constants and $k$-factors in Ref.~\cite{Chung:2009qs}.  As a result of this ``superequilibrium" condition, one can define common charge densities for the various particles and their corresponding superpartners: $U_i$ for right-handed up-type (s)quarks, $D_i$ for left-handed down-type (s)quarks, $Q_i$ for left-handed (s)quarks, $H$ for the combined Higgs-Higgsino density, $R_i$ for the right handed (s)leptons, and $L_i$ for left-handed (s)leptons (here $i$ is a generational index).  We also use the notation $Q\equiv Q_3$, $T\equiv U_3$, $B\equiv D_3$, $L\equiv L_3$, and $R\equiv R_3$, while the $k$-factors for these densities are defined by $k_I=k_i+k_{\tilde{i}}$.  In terms of these definitions, the fermionic part of the density $I$ (the quantity entering the weak sphaleron equation for the LH densities) is given by \begin{equation} n_{i}=\frac{k_{i}}{k_I} I \end{equation} and the LH fermionic charge density $n_L$ is \begin{equation} \label{eq:nL} n_L=\sum _{i=1}^3 \frac{k_{q_i}}{k_{Q_i}} Q_i + \sum _{i=1}^3 \frac{k_{l_i}}{k_{L_i}} L_i. \end{equation}  Two more observations allow us to reduce the number of equations needed to solve for the various densities.  First, since weak sphalerons are decoupled from the Boltzmann equations, baryon and lepton number are approximately locally conserved, so that the sum of all the densities vanishes at a given spacetime point.  Second, since the first and second generation Yukawa couplings are negligible compared to corresponding couplings for the third generation, a first and second generation quark charge can arise only through strong sphalerons, and thus all corresponding charges will be produced in equal number, i.e. $Q_1=Q_2=-2U_1=-2U_2=-2D_1=-2D_2$.  Combined, these two relations imply $B=-(T+Q)$ so that we can eliminate the set of equations governing the $B$ density as well as all of the other first and second generation (s)quark densities besides $Q_1$.

Given the above assumptions, the relevant set of Boltzmann equations to consider are: \begin{equation} \begin{aligned}  \label{QBE1} \partial_{\mu} Q^{\mu} = &-\Gamma_{yt}\left(\frac{Q}{k_Q}-\frac{T}{k_T}+\frac{H}{k_H} \right) -\Gamma_{yb}\left(\frac{Q}{k_Q}+\frac{T+Q}{k_B}-\frac{H}{k_H} \right)  \\ &-\Gamma_{mt} \left(\frac{Q}{k_Q} - \frac{T}{k_T} \right) -\Gamma_{mb}\left(\frac{Q}{k_Q}+\frac{T+Q}{k_B}\right) - S_{\tilde{t}}^{CPV}-S_{\tilde{b}}^{CPV} \\  &- 2\Gamma_{ss} \left( 2\frac{Q}{k_Q}-\frac{T}{k_T}+\frac{Q+T}{k_B} + \frac{1}{2} \sum_{i=1}^2 \left[4 \frac{1}{k_{Q_i}}+\frac{1}{k_{U_i}}+\frac{1}{k_{D_i}}\right] Q_1 \right) \end{aligned} \end{equation} \begin{equation} \begin{aligned} \label{QBE2} \partial_{\mu}T^{\mu} = \hspace{.1cm} &\Gamma_{yt}\left(\frac{Q}{k_Q}-\frac{T}{k_T}+\frac{H}{k_H} \right)  +\Gamma_{mt}\left(\frac{Q}{k_Q} - \frac{T}{k_T} \right) +S_{\tilde{t}}^{CPV} \\ &+\Gamma_{ss} \left( 2\frac{Q}{k_Q}-\frac{T}{k_T}+\frac{Q+T}{k_B} + \frac{1}{2} \sum_{i=1}^2 \left[4 \frac{1}{k_{Q_i}}+\frac{1}{k_{U_i}}+\frac{1}{k_{D_i}}\right] Q_1 \right) \end{aligned} \end{equation} \begin{equation} \begin{aligned} \label{QBE3} \partial_{\mu}Q_1^{\mu} = &-2\Gamma_{ss}\left( 2\frac{Q}{k_Q}-\frac{T}{k_T}+\frac{Q+T}{k_B} + \frac{1}{2} \sum_{i=1}^2 \left[4 \frac{1}{k_{Q_i}}+\frac{1}{k_{U_i}}+\frac{1}{k_{D_i}}\right] Q_1 \right) \end{aligned}\end{equation} \begin{equation} \begin{aligned} \label{QBE4} \partial_{\mu} H^{\mu} = &-\Gamma_{yt}\left(\frac{Q}{k_Q}-\frac{T}{k_T}+\frac{H}{k_H} \right) +\Gamma_{yb}\left(\frac{Q}{k_Q}+\frac{T+Q}{k_B}-\frac{H}{k_H} \right) \\ &+\Gamma_{y\tau}\left(\frac{L}{k_L}-\frac{R}{k_R}-\frac{H}{k_H}\right) -\Gamma_h \frac{H}{k_H} \end{aligned} \end{equation} \begin{equation}\begin{aligned} \partial_{\mu}L^{\mu}=-\Gamma_{y\tau} \left(\frac{L}{k_L}-\frac{R}{k_R}-\frac{H}{k_H}\right) - \Gamma_{m\tau}\left(\frac{L}{k_L}-\frac{R}{k_R}\right)-S_{\tau}^{CPV} \end{aligned} \label{QBE5} \end{equation} \begin{equation}\begin{aligned} \label{eq:QBE_last} \partial_{\mu}R^{\mu}=\Gamma_{y\tau} \left(\frac{L}{k_L}-\frac{R}{k_R}-\frac{H}{k_H}\right) +\Gamma_{m\tau}\left(\frac{L}{k_L}-\frac{R}{k_R}\right)+S_{\tau}^{CPV} \end{aligned}\end{equation}   We solve these equations in the so-called diffusion approximation, in which one introduces a diffusion constant for each species $D_i$ and assumes $\mathbf{j}_i=D_i \nabla n_i$.  The diffusion constants we use are those found in Ref.~\cite{Chung:2009qs}: $D_Q=D_T=D_{Q_i}\simeq 6/T$, $D_H\simeq100/T$, $D_L\simeq100/T$, $D_R\simeq380/T$ where $T$ is the EWPT temperature, assumed to be $100$ GeV.  Note that the left- and right-handed (s)lepton diffusion constants are different; this is because of the $SU(2)$ interactions active in the plasma for LH-densities.  We neglect this difference for the (s)quark diffusion constants since $D_{Q,T,Q_i}$ are determined primarily by $SU(3)$ interactions which are non-chiral. 

With our framework in place, we can now compute the various sources and rates based on previous work in Refs.~\cite{Lee:2004we, Chung:2008aya, Lepton_Mediated, Supergauge, Including_Yukawa, Chung:2009qs} for the stop, sbottom, and stau cases.  We assume the transition temperature $T_c=100$ GeV throughout.

\section{Stop sources}\label{sec:stops}
\begin{figure*}[!t]
\mbox{\hspace*{-1.2cm}\includegraphics[width=0.5\textwidth,clip]{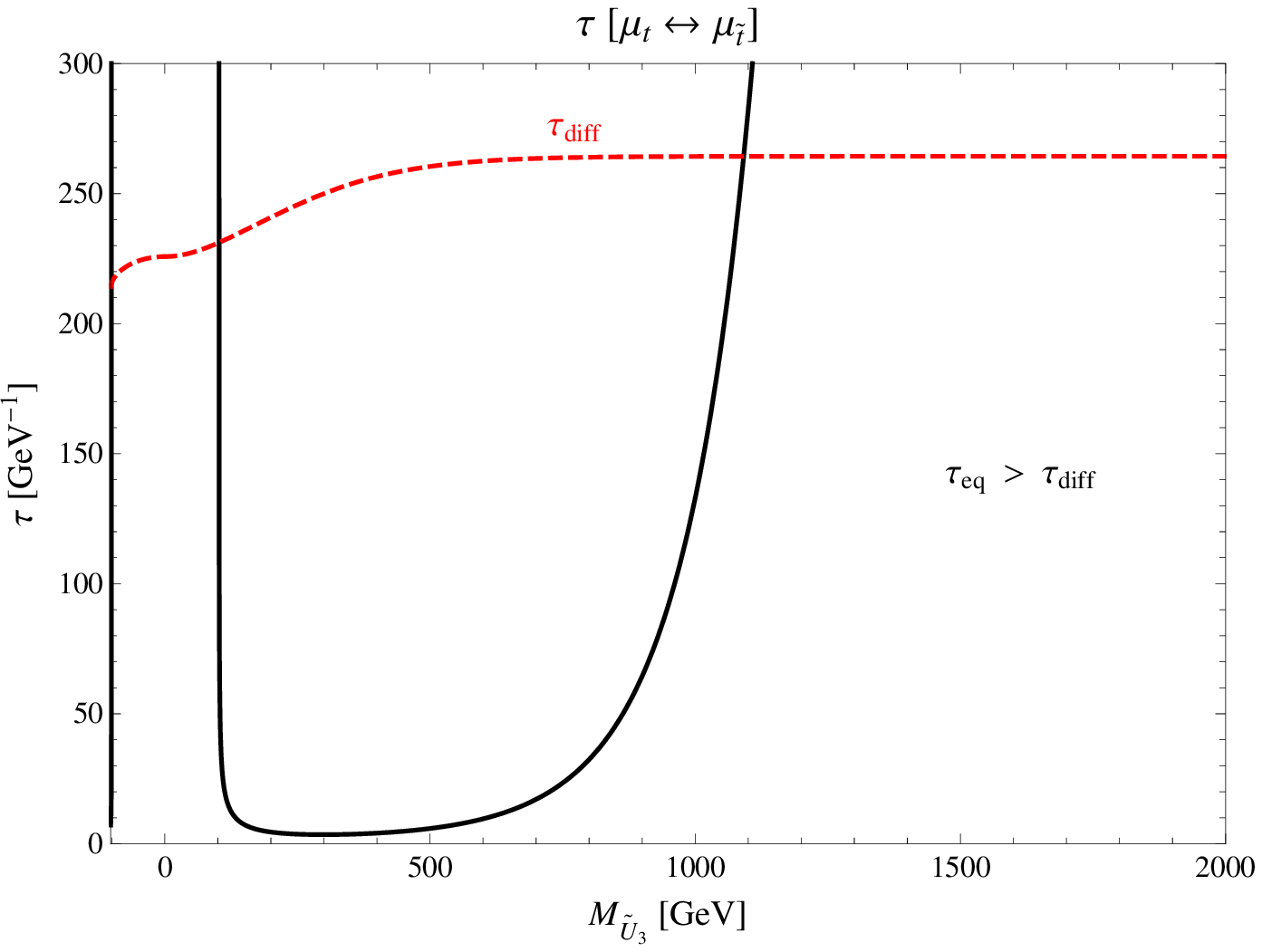}\hspace{.2cm} \includegraphics[width=0.5\textwidth,clip]{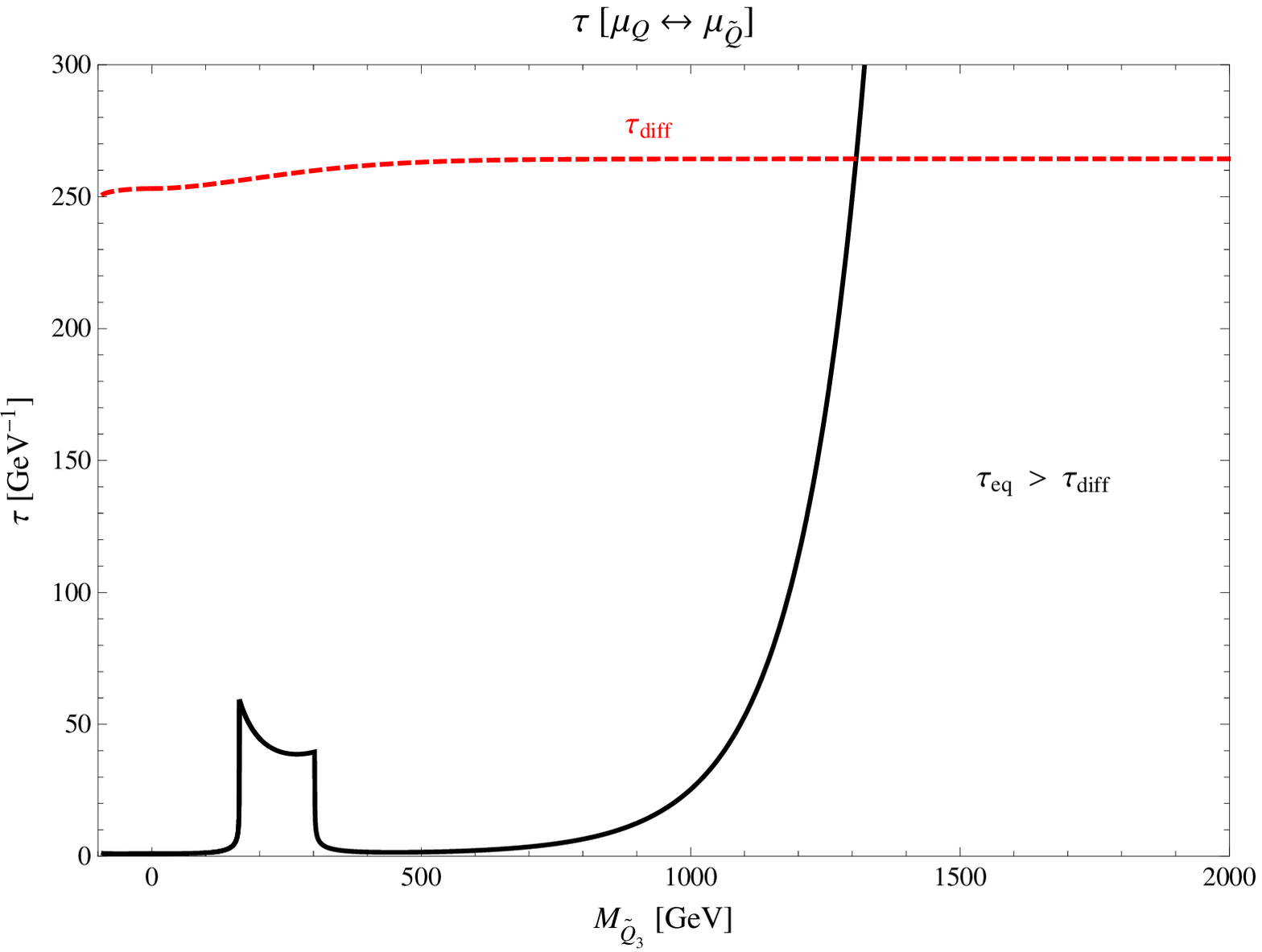}}
\caption{\label{fig:supergauge_rates}\it\small Supergauge equilibration time scales for the RH (s)tops (Left) and LH (s)quarks (Right), where $M_{\tilde{Q}_3}$  ($M_{\tilde{U}_3}$) $=1000$ GeV in computing the RH (LH) stop rates and $M_1=100$ GeV, $M_2=200$ GeV.  Also shown is the diffusion time-scale $\tau_{\rm diff}$ in both cases.  The superequilibrium timescale is longer than $\tau_{\rm diff}$ only in kinematically forbidden regions and for heavy squarks, where the baryon asymmetry is suppressed.}
\end{figure*}

We begin our study by focusing on the scenario in which the observed baryon asymmetry originated primarily from the stop sector.  When the stops scatter off of the spacetime-dependent Higgs vevs in the bubble wall, the CP-violating phase $\phi_t$ arising between the tri-scalar coupling $A_t$ and $\mu$ results in a non-zero expectation value of the current density $\tilde{t}^{\mu}_R$ in and in front of the wall, governed by \begin{equation} \partial_{\mu}\tilde{t}^{\mu}_R(x)=S_{\tilde{t}_R}(x,\left\{n_i\right\}). \end{equation} Here $S_{\tilde{t}_R}(x,\left\{n_i\right\})$ contains both the CP-violating source term as well as the CP-conserving chirality changing rates that also arise from stop scattering off of the Higgs vevs in the plasma, Yukawa interactions and strong sphaleron rates.  To obtain the Boltzmann equations for the stop case as in Eqs.~(\ref{QBE1}-\ref{eq:QBE_last}), one must verify that the supergauge interactions governing the various particle and sparticle densities are in fact in equilibrium for the range of parameters we consider.  Since we will vary the soft breaking masses of both stops, one should be concerned that in some regions of the parameter space, the supergauge interactions involving $\tilde{t}_R$ and $\tilde{t}_L$ will be slow (since these rates are Boltzmann suppressed) or kinematically forbidden, since these rates arise from three-body interactions of the (s)quarks with gauginos.  We plot the supergauge equilibration time scale in Fig.~\ref{fig:supergauge_rates} along with $\tau_{\rm diff}$ for comparison (note that the kinematically forbidden region depends on the precise choice for the gaugino masses).  While the supergauge rate $\Gamma_V^{\tilde{t}t}$ is kinematically forbidden for very light RH stops, $\Gamma_V^{\tilde{q}q}$ is nowhere forbidden.  This is because the latter is a sum of both $\tilde{W}$ and $\tilde{B}$ contributions which are disallowed for different $M_{\tilde{Q}_3}$, and so when $\tilde{W}$ interactions are disallowed, $\tilde{B}$ interactions can still be active and vice versa (again, this depends on the details of the gaugino masses).  Everywhere else both LH and RH rates are quite fast compared to $\tau_{\rm diff}$. The only other exception is the region corresponding to heavy squarks, where the baryon asymmetry is also expected to be suppressed.  Since $\tau_{eq}\ll \tau_{\rm diff}$ for most of the parameter space relevant for EWB  we work under the simplifying approximation that stop-top supergauge equilibrium holds in all regions of interest when computing $Y_B$.  Additionally, there are supergauge rates involving the other charge densities occurring in Eqs.~(\ref{QBE1}-\ref{eq:QBE_last}): we have verified that the corresponding rates for Higgs and higgsino densities are also fast compared to $\tau_{\rm diff}$ for our choices of parameters, detailed below.  The supergauge rates for the heavy squarks we consider are suppressed, and their equations decouple from the full set in Ref.~\cite{Chung:2009qs}.  As a result,  the density $Q_1$ consists entirely of fermions, $Q_1=q_1$. 

%%%%%%%%%%%%%%%%%%%%%%%%%%%%%%%%%%%%%%%%%%%%%%%%%%%
%%%%%%%%%%%%%%%%%%%%%%%%%%%%%%%%%%%%%%%%%%%%%%%%%%%

In computing the baryon asymmetry we use the form of the sources computed in Ref.~\cite{Lee:2004we} and related work, which exhibit resonant behavior for nearly degenerate RH- and LH- stop masses.  We quote the form of the stop CP-violating source $S_{\tilde{t}}^{CPV}$  here, to allow straightforward generalizations to the cases of sbottom and stau sources in the following sections: 
\begin{equation} 
\label{eq:stop_source} \begin{aligned} S_{\tilde{t}}^{CPV}(x)=&\frac{N_Cy_t^2}{2\pi^2}\operatorname{Im}(\mu A_t)v^2(x)\dot{\beta}(x) \\ & \times \int_0^{\infty} \frac{dk k^2}{\omega_R \omega_L}  \operatorname{Im}\left[\frac{n_B(\mathcal{E}^*_R)-n_B(\mathcal{E}_L)}{(\mathcal{E}_L-\mathcal{E}_R^*)^2}+\frac{n_B(\mathcal{E}_R)+n_B(\mathcal{E}_L)}{(\mathcal{E}_L+\mathcal{E}_R)^2} \right] \end{aligned} 
\end{equation}  
The various quantities involved are given by \begin{align} \label{eq:definitions} \mathcal{E}_{L,R}&=\omega_{L,R}-i\Gamma_{L,R}\\ \omega_{L,R}&=\sqrt{\left|\mathbf{k}\right|^2+m_{\tilde{t}_{L,R}}^2} \\ h_B(x)&=-\frac{e^{x/T}}{(e^{x/T}-1)^2}\\ n_B(x)&=\frac{1}{e^{x/T}-1} \end{align} where $\Gamma_i$ are the thermal widths of the stops which are $\mathcal{O}(10^{-1}T)$, $v_{u,d}$ are the spacetime-dependent Higgs vevs, $v^2=v_u^2+v_d^2$, $\tan \beta=v_u/v_d$, $y_t$ is the top Yukawa coupling, $T$ is the EWPT temperature, $N_C$ is the number of colors, and $m_{\tilde{t}_{R,L}}$ are the effective stop masses given in terms of the corresponding soft breaking and thermal masses by $m^2_{\tilde{t}_{L,R}}\equiv M^2_{\tilde{Q}_3,\tilde{U}_3}+M^2_{T;L,R}$.  The dependence of Eq.~(\ref{eq:stop_source}) on the CP-violating phase $\phi_t$ is apparent.  Also, the CP-violating source (and the chirality-changing CP-conserving rates) are manifestly spacetime-dependent, as they are proportional to the Higgs vevs.  We use a simplified step-function profile for these rates and sources, deferring a careful treatment of the bubble profile to future study. Note also that we have omitted a temperature-independent contribution to the numerator of the second term in Eq.~(\ref{eq:stop_source}) that appears in the corresponding expression in Ref.~~\cite{Lee:2004we}.  The current density from which the CP-violating source is derived must be properly normal-ordered through a subtraction of the zero-temperature matrix element. Implementing this normal ordering effectively removes the temperature-independent contribution to the numerator\footnote{We thank C. Lee for this observation and T. Liu for highlighting this issue in an earlier version of this work.}.

In addition to the CP-violating source Eq.~(\ref{eq:stop_source}), we use the form of the relaxation, Yukawa, triscalar, and strong sphaleron rates worked out in Ref.~\cite{Chung:2009qs}; we do not reproduce them here for brevity.  Since we are interested in only the stop CP-violating source contribution to the BAU, we take the RH sbottom and RH, LH stau masses to be heavy which allows us to neglect  the (s)bottom, (s)tau Yukawa rates - with heavy superpartners, only SM-like Yukawa interactions contribute to these rates, resulting in $\Gamma_{yb,\tau} \ll \Gamma_{yt}$, $\Gamma_{ss}$ in virtually all of the parameter space we consider.  With this choice of spectrum we can also neglect the CP-conserving chirality changing rates $\Gamma_{mb}$ for the (s)bottoms, which are suppressed by factors of $\left( \frac{y_b}{y_t}\right)^2 \cot^2 \beta$ with respect to $\Gamma_{mt}$ \cite{Lepton_Mediated}.  With these simplifications, the only source for $B$ charge density are strong sphalerons, implying that $B=-(Q+T)=-\frac{1}{2}Q_1$ and consequently simplifying Eq.~(\ref{eq:nL}) to \begin{equation} \label{eq:nL_1} n_L=\frac{k_q}{k_Q}Q+4(Q+T).\end{equation}  Due to the relation between $B$ and $Q_1$ and the decoupling of the (s)leptons which occurs when neglecting their Yukawa couplings, the full set of Boltzmann equations reduces  to Eqs.~(\ref{QBE1}), (\ref{QBE2}), and (\ref{QBE4}) with the replacements $\Gamma_{yb,y\tau}$, $\Gamma_{mb}\rightarrow 0$, $Q_1\rightarrow 2(Q+T)$, coinciding with the set described in Ref.~\cite{Lee:2004we} and which we use in our numerical computation of the BAU.

There are several uncertainties built into our computation of the baryon asymmetry.  In addition to those arising from the vev-insertion approximation, theoretical uncertainties in several other parameters associated with the phase transition such as the bubble wall thickness, velocity, and variation of the Higgs vevs $\Delta\beta$ can introduce $\mathcal{O}(100$ GeV) uncertainties in the curves of constant baryon density, similarly to the case of higgsino/gaugino sources (see \cite{Kozaczuk:2011vr} and references therein).  For concreteness, we consider conservative values for the wall velocity, $v_w=5/100$, and thickness, $L_w=5/T_c$.  Additionally, non-resonant sources such as those computed in Refs.~\cite{Carena:2002ss} and related work yield results for the BAU that can differ significantly from the values computed using the vev-insertion approximation, especially away from the resonance. To take into account the uncertainties associated with a precise calculation of the BAU, we show on selected plots curves corresponding to $10 \times Y_{Obs}$, as a rough upper bound on the stop source scenario, as well as curves of $0.1 \times Y_{Obs}$ as a more conservative estimate of the BAU.  As we will see, our conclusions hold across this wide range of uncertainty.

\subsection{Parameter Space }\label{sec:params}
The baryon asymmetry produced by stop sources depend on the masses $m_{\tilde{t}_{L,R}}$, which are temperature dependent.  Since the thermal masses are constant at a given temperature, we can equivalently investigate the potential of stop sources to produce the observed BAU by varying the values of the LH and RH stop SUSY breaking soft masses, $M_{\tilde{Q}_3,\tilde{U}_3}$.  We vary $M_{\tilde{U}_3}$ over the range  $-100^2$ GeV$^2\le M^2_{\tilde{U}_3}\le 2000^2$ GeV$^2$, which includes the so-called ``light stop scenario" for $M^2_{\tilde{U}_3}<0$ (and multi-TeV $M_{\tilde{Q}_3}$), a region of the parameter space where the light RH stop provides the strongly first order phase transition needed for successful baryogenesis.  We stress that away from negative values of $M^2_{\tilde{U}_3}$, some other mechanism is needed to generate a strongly first order phase transition.  Several such mechanisms have been proposed \cite{Heavy_Stops, Profumo:2007wc} that are decoupled from the spectrum required for EWB and from the physical processes of interest for the present discussion.  Thus, in evaluating the potential of stops, sbottoms, and staus for EWB, we consider only the strength of the CP-violation in each case and assume a strongly first-order EWPT generated by one of these other mechanisms.  For the LH stops, we vary $M_{\tilde{Q}_3}$ over the range  $100$ GeV $\le M_{\tilde{Q}_3}\le 4000$ GeV.  Within these mass ranges there are regions where the choice of soft mass leads to negative or zero mass squared for the lightest stop at $T=0$ for various values of the triscalar coupling and $\mu$; we indicate these regions (along with more stringent constraints on stop masses from direct searches discussed in Sec.~\ref{sec:constraints}) on all of our plots.     For the stops, we focus on $\tan \beta=10$.  One should note that larger $\tan\beta$ yields larger SM-like Higgs masses along with more stringent EDM constraints.

There are several other parameters whose values need to be fixed in order to calculate the BAU.  We choose values for these parameters conservatively, bearing in mind the various constraints from Higgs mass measurements, stop searches, and EDM search null results as well as theoretical considerations such as the avoidance of color and charge-breaking vacua.  
In computing the baryon asymmetry, we take $m_A=200$ GeV.  For larger $m_A$, the baryon asymmetry is reduced due to the dependence of $\Delta \beta$ on $m_A$, which scales as $\Delta \beta \sim 1/m_A^2$ \cite{Carena:2002ss}.  The gaugino soft masses are taken to be real, with $M_1=80$ GeV, $M_2=250$ GeV to ensure a light neutralino $\chi^0_i$ as the lightest supersymmetric particle (LSP) while other gauginos are rather heavy.  For the scenario we consider here the resulting baryon asymmetry and Higgs mass constraints do not depend sensitively on $M_1$, $M_2$.  The gluino soft mass is largely decoupled from the phenomenology relevant here, and was set to $M_3=10$ TeV.  For the higgsino mass parameter $\mu$ (which we take to be real, so that $\phi_t$ arises only from the phase in $A_t$), we choose $\mu=200$ GeV, 1000 GeV to illustrate the behavior of the baryon asymmetry and the various constraints in these cases.  Small values of $\mu$ suppress the BAU (c.f. Eq.~(\ref{eq:stop_source})), while large values can make the zero-temperature physical stop masses squared negative by making the off-diagonal components of the mixing matrix large, as well as strengthen the various EDM constraints.  Similarly, we vary the magnitude of the trilinear scalar coupling $\left|A_t\right|=100$ GeV, $250$ GeV, $1000$ GeV; larger values of $\left|A_t\right|$ also result in larger exclusions from EDM constraints.  We typically consider the case of maximal CP-violating phase $\phi_{t}=\pi/2$ to show the maximal extent of the EWB-compatible parameter space.  We rely on this phase to produce all of the baryon asymmetry, setting all other CP-violating phases $\phi_i=0$ to isolate the contribution from the stop sources to the BAU.  Otherwise, non-stop sources such as those arising from gaugino-higgsino-vev interactions will further contribute to the baryon asymmetry and there will be additional contributions to the EDM constraints.  Finally, all other triscalar couplings are taken to be zero, and all other sfermions in our analysis are taken to be heavy, $m_{sf}=10$ TeV.  As shown in Ref.~\cite{Chung:2009qs}, this effectively decouples them from the network of transport equations, since superequilibrium and Yukawa rates that can transfer charge density between SM particles and their superpartners vanish for any of the masses much larger than the temperature.  As a result, the densities $\left\{I\right\}$ appearing in the transport equations for these sfermions (e.g. $Q_1$) correspond entirely to an SM particle charge density, $k_I=k_i$.    

\begin{figure*}[!t]
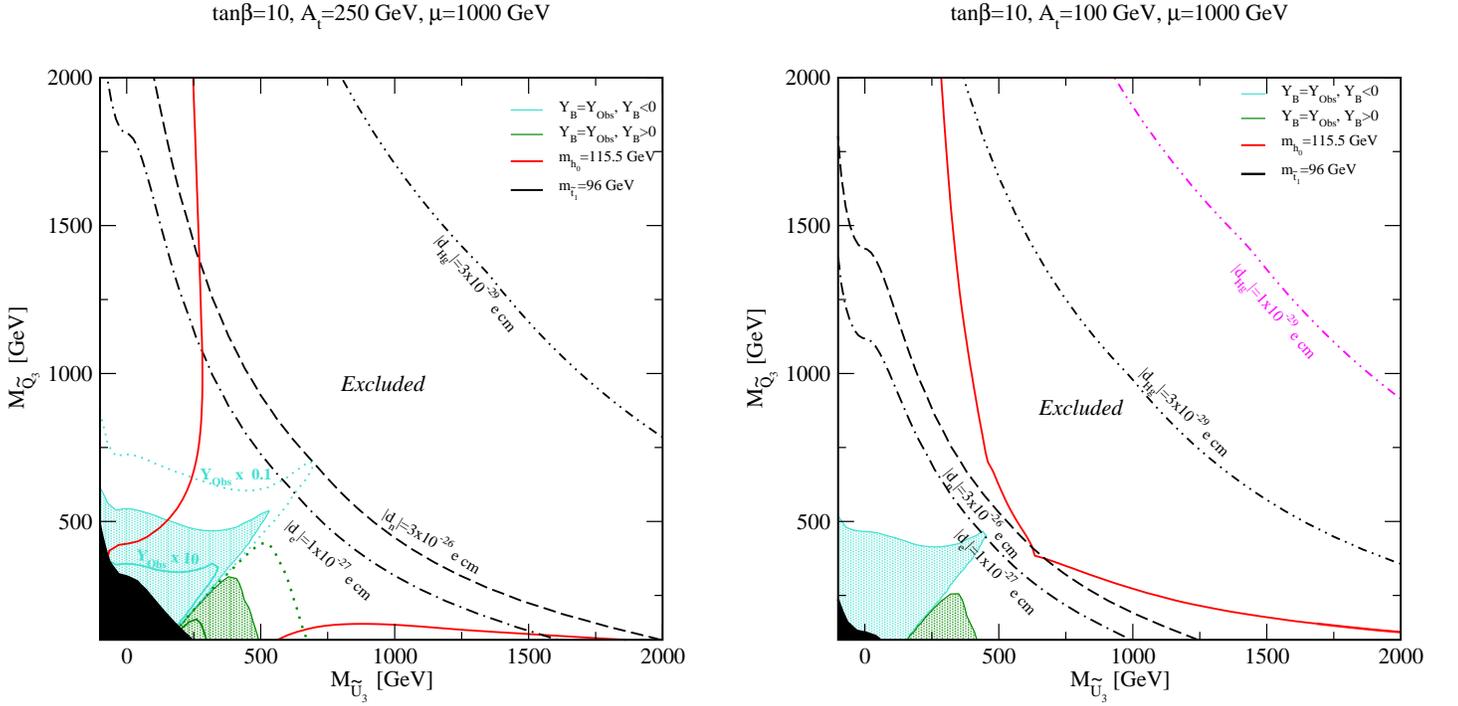

\mbox{\hspace*{-1.2cm}\includegraphics[width=0.55\textwidth,clip]{definitive_plots/tanb_At_mu_10_250_1000.eps}\qquad\includegraphics[width=0.55\textwidth,clip]{definitive_plots/tanb_At_mu_10_100_1000.eps}}
\caption{\label{fig:stop_sources1}\it\small Regions of the stop soft supersymmetry breaking mass parameter space consistent with the observed value of the baryon asymmetry resulting from stop sources for $\mu=1000$ GeV, $\left|A_t\right|=250$ GeV (Left) and $\left|A_t\right|=100$ GeV (Right).  Regions shaded blue (green) correspond to $Y_B\geq Y_{Obs}$ with $Y_B<0$ ($Y_B>0$) for maximal CP-violating phase.  The dotted blue contour on the left marks the region that would be consistent with stop-sourced EWB if the vev-insertion approximation had underestimated $Y_B$ by a factor of 10 (we omit this curve in subsequent plots).  On the left we also show, by the darker shaded regions, the parameter space compatible with $10\times$ the observed BAU, i.e. the allowed regions if the vev-insertion approximation overestimated $Y_B$ by a factor of 10.  Black shaded regions are excluded by stop mass direct searches; regions to the left of the thick red line are excluded by LEP Higgs mass bounds in both cases.  Current constraints on the electron, neutron, and $^{199}$Hg EDMs  are represented by the black dashed-dot, dashed, and dashed-double-dot lines, respectively, with regions to the left of each line ruled out by null results; the projected future reaches for $d_e$, $d_n$, and $d_{Hg}$ measurements are shown in magenta (where applicable).  In both cases here, both the $d_e$ and $d_n$ future sensitivities lie above the plane shown.  For the $\left|A_t\right|=250$ GeV case, the Mercury EDM future sensitivity also lies above the plane shown.}
\end{figure*}

\begin{figure*}[!t]
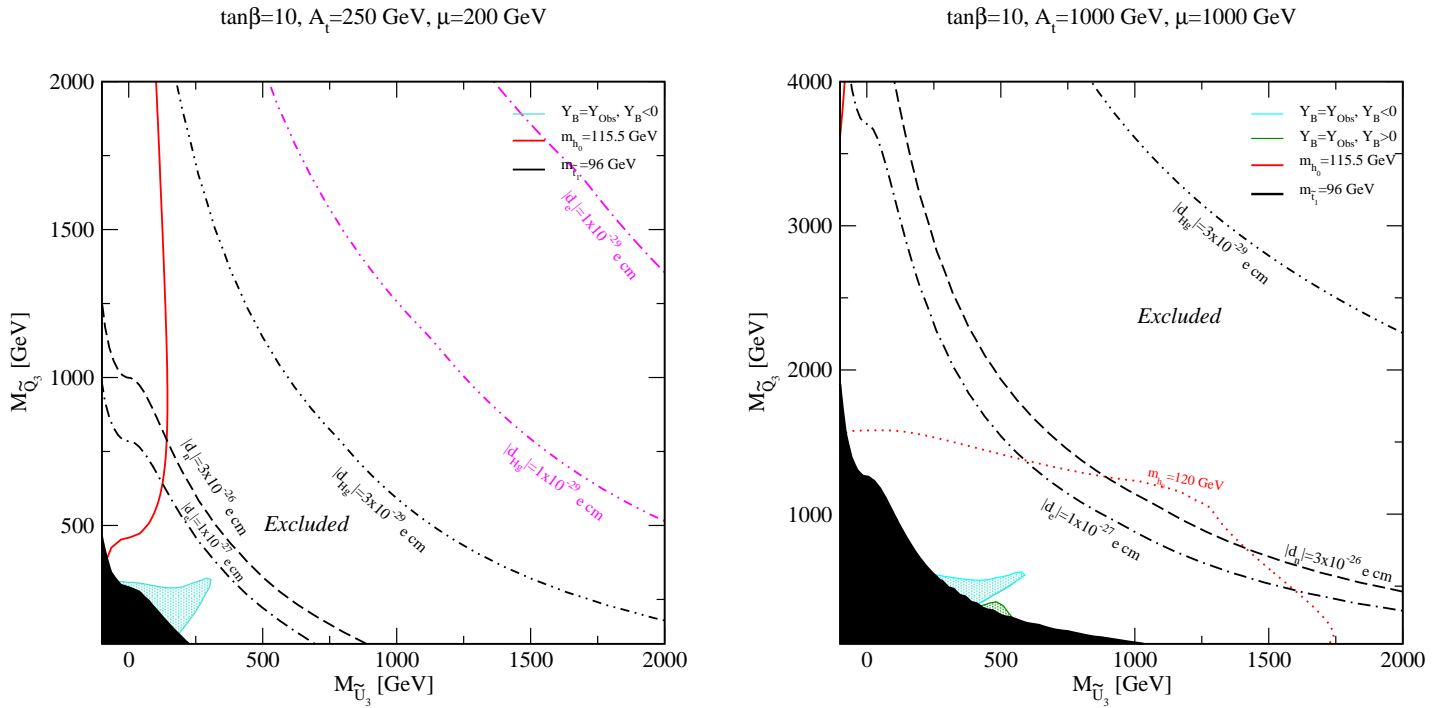

\mbox{\hspace*{-1.2cm}\includegraphics[width=0.55\textwidth,clip]{definitive_plots/tanb_At_mu_10_250_200.eps}\qquad\includegraphics[width=0.55\textwidth,clip]{definitive_plots/tanb_At_mu_10_1000_1000.eps}}%\qquad\includegraphics[width=0.55\textwidth,clip]{definitive_plots/tb_At_mu_10_250_1000_plotting.eps}}
\caption{\label{fig:stop_sources2}\it\small  Same as Fig.~\ref{fig:stop_sources1}, but for $\left|A_t\right|=250$ GeV, $\mu=200$ GeV (Left) and $\left|A_t\right|=1000$ GeV, $\mu=1000$ GeV (Right).  For $\left|A_t\right|=250$ GeV, the $Y_B>0$ curve falls beneath the black shaded region and future measurements of the neuron EDM are expected to probe all parameter space shown.  For $\left|A_t\right|=1000$ GeV, the expected reaches of $d_e$, $d_n$, and $d_{Hg}$ future measurements lie above the plane shown here.}
\end{figure*}

Using this spectrum, we calculate the baryon asymmetry generated by stop scattering off of the bubble wall and outline regions of the stop mass parameter space suitable for successful EWB in Figs.~\ref{fig:stop_sources1}-\ref{fig:stop_sources2}; regions consistent with the observed value of $Y_B$ are shaded.  The contours shown correspond to maximal CP-violating phase $\sin\phi_t$, while for smaller phase the baryon asymmetry is suppressed as are the EDM constraints.  Decreasing $|\sin\phi_t|$ does not open up any additional parameter space for EWB.  Several important features of the sources are shown in Figs.~\ref{fig:stop_sources1}-\ref{fig:stop_sources2}.  From Eqs.~\ref{eq:stop_source}, \ref{eq:definitions}, the CP-violating source is resonant for $m_{\tilde{t}_R}\sim m_{\tilde{t}_L}$.  This manifests itself as a resonance for $M^2_{\tilde{U}_3}\sim M^2_{\tilde{Q}_3} + (M^2_{T;L}-M^2_{T;R})\approx M^2_{\tilde{Q}_3} $ in the parameter space as shown.  

Also, there is an increase in the generated BAU for $M^2_{\tilde{U}_3}<0$.  This feature arises far from the resonance \footnote{We caution the reader that away from the resonance, there may also be non-resonant contributions to the sources \cite{Carena:2002ss} which we do not consider here.} but is straightforward to understand from the form of the CP-violating stop sources.  The quantities $w_{R,L}$, $\mathcal{E}_{R,L}$ entering into $S^{CPV}_{\tilde{t}}$ depend on the physical masses $m^2_{\tilde{t}_{R,L}}= M^2_{\tilde{Q}_3,\tilde{U}_3}+M^2_{T;R,L}$ and for $M^2_{\tilde{U}_3}\rightarrow -M^2_{T;R}$, the physical mass $m_{\tilde{t}_{R}}\rightarrow 0$.  In this regime, the Boltzmann distributions in the integrand for $S_{\tilde{t}}$ begin to diverge for $k=0$, $n_B(k)\rightarrow1/ \left(e^{\left|k\right|-i\Gamma}-1\right)$ which corresponds physically to the abundance of nearly massless squarks produced in the thermal bath.  We emphasize that $M_{\tilde{U}_3}^2<0$ does not result in a tachyonic stop in the unbroken phase as long as $M_{\tilde{U}_3}^2>-M^2_{T;R}$ (the thermal masses are $\mathcal{O}(100$ GeV) so this is not an issue in the parameter space we consider).  On the other hand, the zero-temperature stop mass eigenstates can turn negative in some of the parameter space; the corresponding regions are of course ruled out by direct searches for stops, corresponding to the black shaded regions in Figs.~\ref{fig:stop_sources1}-\ref{fig:stop_sources2}.

Finally, we find that there are regions for which the produced baryon asymmetry switches sign.  This effect arises due to the competition between $Q$ and $T$ densities in the expression for $n_L$, Eq.~(\ref{eq:nL_1}); since $Q$ and $T$ densities carry opposite sign, when $M_{\tilde{Q}_3}$ is small, $k_q/k_Q$ decreases and the $T$ contribution can win out and drive $n_L\geq0$.  In Figs.~\ref{fig:stop_sources1}-\ref{fig:stop_sources2}, regions with $Y_B>0$ are shaded green, while regions for which $Y_B<0$ are shaded blue. Since at present the phase $\phi_t$ is not experimentally constrained, either region can lead to the appropriate overall sign for the baryon asymmetry through an appropriate choice of  $\phi_t$.

%%%%%%%%%%%%%%%%%%%%%%%%%%%%%%%%%%%%%%%%%%%%%%%%%%%
%%%%%%%%%%%%%%%%%%%%%%%%%%%%%%%%%%%%%%%%%%%%%%%%%%%

%%%%%%%%%%%%%%%%%%%%%%%%%%%%%%%%%%%%%%%%%%%%%%%%%%%
%%%%%%%%%%%%%%%%%%%%%%%%%%%%%%%%%%%%%%%%%%%%%%%%%%%

\subsection{Stop and Higgs Mass Constraints}\label{sec:constraints}

Having calculated the BAU resulting from stop sources, one should ask how the stop mass parameter space consistent with successful EWB confronts various other phenomenological constraints.  We consider three types of constraints on our EWB scenario: stop mass constraints from collider searches, Higgs mass bounds, and electric dipole moment search null results.   

Zero-temperature stop masses have been constrained by direct searches for superpartners  at LEP and the LHC (for particle spectra relevant here) to be $m_{\tilde{t}_1}> 96$ GeV \cite{PDG}.  This lower bound arises from considering stop decays to the LSP, typically assumed to be the lightest neutralino.  With our choice of gaugino masses, the lightest stop $\tilde{t}_1$ is heavier than $\chi_1^0$ in all of the parameter space so that this lower bound on $m_{\tilde{t}_1}$ is applicable.  There are several specific cases in which the stop masses might be more tightly constrained, however for generality we consider this lower bound for our scenario.  We calculate the physical $T=0$ stop masses using \texttt{FeynHiggs} \cite{FeynHiggs} for the choices of parameters discussed above and indicate $m_{\tilde{t}_1}\leq 96$ GeV on our plots by the black shaded region.  Increasing $\left|A_t\right|$ leads to larger regions of parameter space for which the lightest stop falls below the lower bound.  This is because the triscalar coupling appears in the off-diagonal entries in the stop mass matrix and large values reduce the value of the smaller eigenvalue. 

The mass of the SM Higgs has been constrained by LEP to be $m_{h_0}>115.5$ GeV \cite{Schael:2006cr,ATLAS:2012ae}.  We use \texttt{FeynHiggs} to calculate the mass of the SM-like Higgs to two-loop order and indicate the LEP bound by a thick red line on our plots.  In addition to the lower bound from LEP, recent preliminary results from both ATLAS and CMS experiments have indicated the possibility of a SM-like Higgs with $m_{h_0}\approx 125$ GeV \cite{higgsclaims}.  However, for light stops and small $\left|A_t\right|$, the corrections to $m_{h_0}$ arising from diagrams with stop loops typically needed to increase the mass of the SM-like Higgs in the MSSM are suppressed, and we find no parameter space consistent with $m_{h_0}=125$ GeV.  For larger $\left|A_t\right|$, the stop loop corrections can be enhanced and the Higgs mass can be pushed up to $m_{h_0}\approx 120$ GeV (which we indicate on the plot corresponding to $\left|A_t\right|=1000$ GeV, $\mu=1000$ GeV with a red dotted line), however we find that $m_{h_0}=125$ GeV is difficult to obtain for our choices of parameters.  We note that additional field content, such as the inclusion of a gauge singlet in the superpotential in e.g. the NMSSM, which may be required to provide a strongly first order phase transition in these scenarios, can result in large contributions to $m_{h_0}$, even at tree level.  Thus, our Higgs mass contours should not be taken as strict exclusions, but as illustrating the tension encountered in the MSSM between light third generation scalars and a heavy SM-like Higgs.

\subsection{EDM Constraints}
CP-odd couplings in the MSSM will generally give rise to electric dipole moments (EDMs) of elementary fermions, nucleons, and neutral atoms. To date, no EDM has been experimentally observed, with the most stringent limits having been obtained on the EDM of the neutral Mercury atom \cite{Griffith:2009zz} ($|d_{H\! g}| < 3\times 10^{-29}e\text{ cm}$),   electron (via the YbF molecule) \cite{Hudson:2011zz} ($|d_e| < 1.05 \times 10^{-27} e\text{ cm}$), and neutron \cite{Baker:2006ts} ($|d_n| < 2.9\times 10^{-26} e \text{ cm}$). The non-observation of these EDMs places powerful constraints upon the strength of the CP-odd sources used in EWB (for a discussion of the constraints relevant to Higgsino-Bino-Wino driven MSSM baryogenesis, see, {\em e.g.}, Refs.~\cite{EWB_and_EDMs,Li:2008kz}) . 
On-going efforts could improve the sensitivity of EDM searches by up to two orders of magnitude (for a review, see, {\em e.g.}, Ref.~\cite{Hewett:2012ns}), suggesting the future possibility of even more stringent constraints or the observation of an EDM with a magnitude consistent with the requirements of MSSM EWB. 

In order to analyze the impact of the present and prospective constraints, we use the program \texttt{CPsuperH} \cite{Ellis:2008zy} to compute the relevant EDMs under different scenarios. In particular, when CP-violation is generated entirely by the phase $\phi_t$, the largest contributions to the relevant EDMs are generated by two-loop graphs that give rise to the Weinberg three-gluon operator ($d_G^C$) as well as  to \lq\lq Barr--Zee" graph contributions to the elementary fermion EDM ($d_f^E$) and quark chromo-EDMs ($d_q^C$). In addition,  four-fermion interactions are generated at one-loop order, though the effects of these operators are typically suppressed.

Before proceeding, we note that there exists considerable theoretical uncertainty in the computation of EDMs of strongly-interacting and many-body systems. In the case of diagmagnetic atoms such as $^{199}$Hg, the dominant contribution to the EDM arises from the nuclear Schiff moment induced by CP-violating nucleon-nucleon interactions. In general, the most important contribution to the latter is a long-range effect arising from single pion exchange, wherein one of the pion-nucleon vertices ($\pi NN$) is CP-odd\footnote{Technically speaking, the interaction is odd under parity and time-reversal.} and the other CP-even. The CP-odd $\pi NN$ interaction can be induced by the Weinberg three-gluon operator, CP-odd four-quark operator, and/or quark chromo-EDM operator, though in the MSSM the latter contribution typically dominates \cite{Pospelov:2005pr}. The computation of the atomic EDM, thus, encounters several sources of theoretical uncertainty: the calculation of the CP-violating $\pi NN$ vertices from the underlying CP-violating interaction; the computation of the nuclear Schiff moment that generally requires a scheme for nuclear model-space truncation; and the corresponding atomic physics computation of the induced EDM. 

\begin{figure}[t]
\centering
\subfigure{
\includegraphics[width=3in]{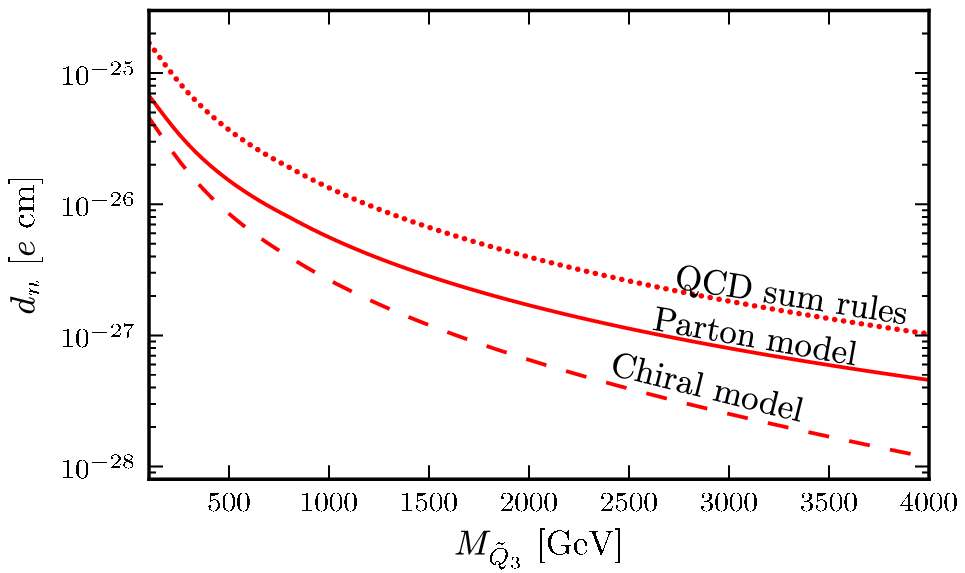}
\label{fig:all_neut}
}
\subfigure{
\includegraphics[width=3in]{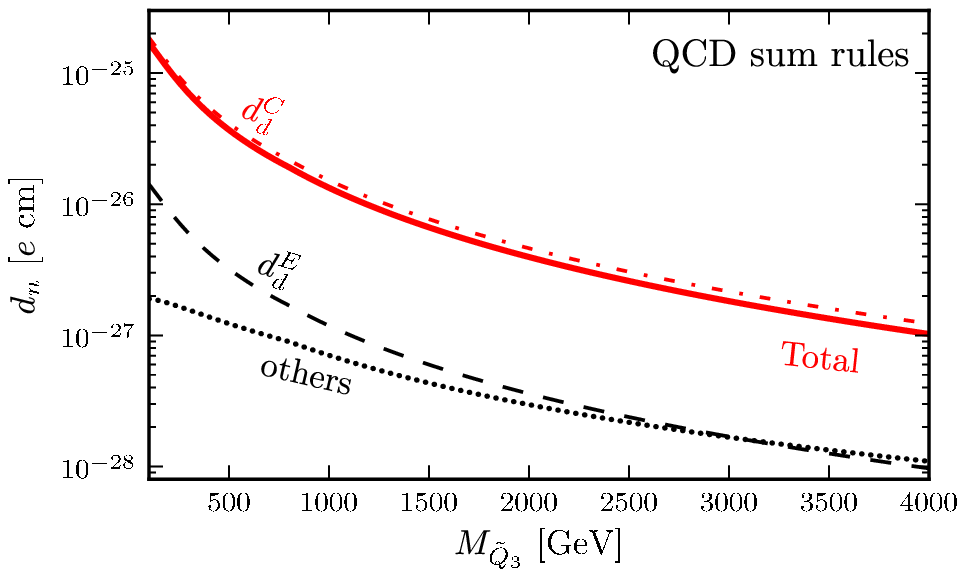}
\label{fig:neut_QCD}
}
\caption{\label{fig:neutEDM}
Neutron EDMs for $M_{\tilde{U}_3} = 800$ GeV, $\tan \beta = 10$, $\mu = 1000$ GeV and $\left|A_t\right| = 250$ GeV. Red denotes negative values. Left: the three independent calculations of the neutron EDM. Right: EDM subcomponents using QCD sum rules. 
By far the largest contribution comes from the down-quark chromo-EDM $d^C_d$, followed by the down-quark EDM $d^E_d$. }
\end{figure}

The computation of the neutron EDM is clearly less susceptible to theoretical uncertainties, as only those associated with hadronic effects enter.  Nonetheless, these uncertainties can be substantial for both the neutron and atomic EDMs. For example, recent work by the authors of Ref.~\cite{Hisano:2012sc} utilizing the QCD sum rule technique suggests that the sensitivity of the neutron EDM to quark EDM and chomo-EDMs may be a factor of five smaller than given by earlier work \cite{Pospelov:2005pr} that provides the basis for the \texttt{CPsuperH} code.  In the case of the nuclear Schiff moment contributions to the $^{199}$Hg EDM, the code has yet to take into account state-of-the-art many body computations \cite{Ban:2010ea} that  imply substantial differences with the many-body calculations using a simplified, schematic nuclear interaction on which the code is based. Consequently, we caution that the precise numerical results associated with the neutron and $^{199}$Hg EDM constraints given below should be taken with a grain of salt (we comment more on the impact of this uncertainty on our results below).  To provide an indication of the kind of theoretical uncertainty one might expect, we show in Fig. \ref{fig:all_neut} computations of the neutron EDM using different approaches as discussed in Ref.~\cite{Ellis:2008zy} (QCD sum rules, the chiral quark model, and parton quark model), though we rely only on the QCD sum-rule technique in our analysis. The QCD sum rule computations tend to give the largest EDM, leading to the strongest constraints. 

\begin{figure}[t]
\centering
\subfigure{
\includegraphics[width=3in]{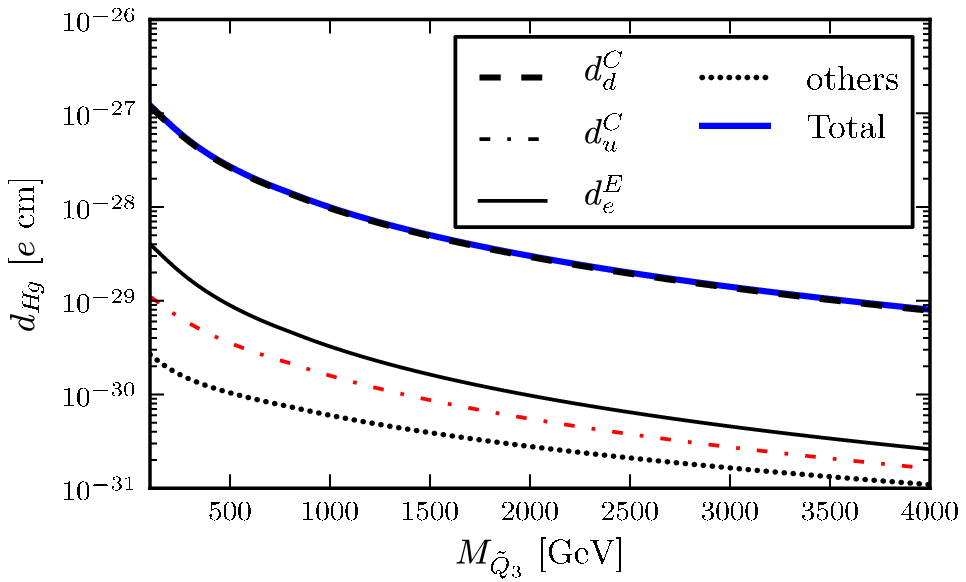}
}
\subfigure{
\includegraphics[width=3in]{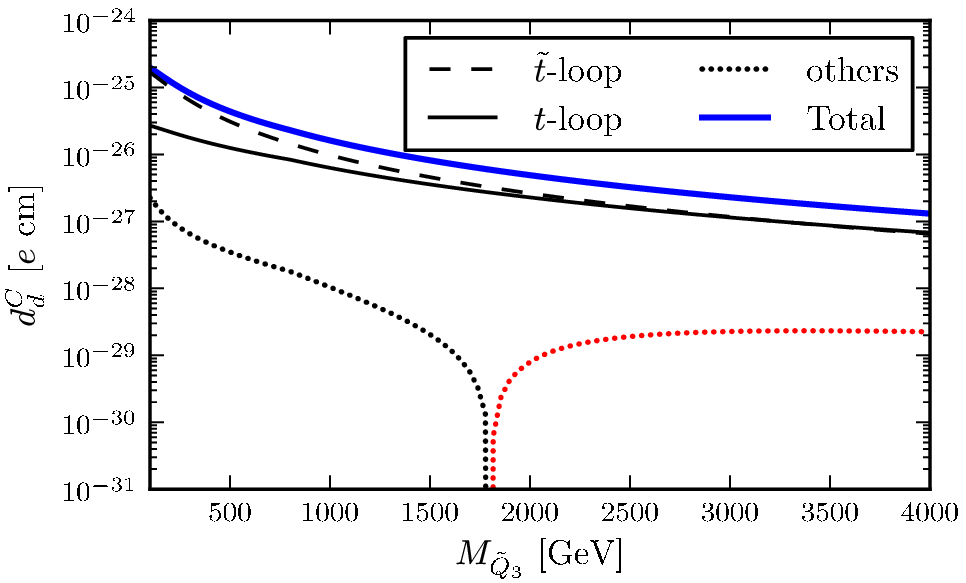}
}
\caption{\label{fig:Hg}
Left: a breakdown of the Mercury EDM, using the same parameters as in figure~\ref{fig:neutEDM}. Almost the entire contribution comes from the down-quark chromo-EDM (multiplied by a constant factor). Right: a further breakdown of the down-quark chromo-EDM.
  }
\end{figure}

With these caveats in mind, we observe that the strongest constraint for stop sources comes from the Mercury EDM, for which, in turn, the quark chromo-EDMs provide the most important contribution as shown in Fig.~\ref{fig:Hg}. We also include the constraint from the electron EDM. For the scenario of interest here, the largest contributions to the $d_e$ EDM come from top and stop loops in Barr--Zee graphs. Note that, like the CP-violating sources, all EDMs are roughly proportional to $|\mu| |A_t|$. Therefore, increasing one of these parameters in order to get a model with sufficiently strong baryogenesis also tends to produce a model that is ruled out by EDM searches.

\subsection{Results}

The various constraints are plotted along with the curves of constant $Y_B$ for different values of $\mu$, $\left|A_t\right|$ in Figs.~\ref{fig:stop_sources1}-\ref{fig:stop_sources2}.  For $\left|A_t\right|=250$ GeV, $\mu=1000$ GeV, one finds that direct search constraints rule out light, nearly-degenerate stop soft-breaking masses, with the bound from LEP on $m_{h_0}$ excluding portions of the parameter space away from the resonance.  Additionally, null results from searches for the electron and neutron EDMs rule out nearly all of the EWB-compatible parameter space except for the tip of the resonance (the expected reach of future EDM experiments are also included in magenta).  The strongest constraint is that arising from searches for the $^{199}$Hg EDM, which rules out all of the viable parameter space, even excluding regions in which the stops produce $1/10$ of the observed BAU.  This remarkable result is due to the stringent limit on the Mercury EDM coupled with the relatively large chromo-EDM contributions of diagrams involving stop loops to $d_{Hg}$.  We find a similar landscape for the $\left|A_t\right|=100$ GeV, $\mu=1000$ GeV case in Fig.~\ref{fig:stop_sources1}; here the BAU is reduced relative to the $\left|A_t\right|=250$ GeV case and direct searches rule out less of the parameter space because of the reduced mixing.  The Higgs mass constraints are stronger due to the smallness of $\left|A_t\right|$ and the LEP bound alone rules out all of the available parameter space for stop-sourced EWB.  But once again the $^{199}$Hg EDM constraint is the most significant, excluding all parameter space in which the stops produce even $1/10$ of the observed BAU.  Moving on to the $\left|A_t\right|=250$ GeV, $\mu=200$ GeV scenario in Fig.~\ref{fig:stop_sources2}, one might imagine that smaller values of $\mu$, $\left|A_t\right|$ may reduce the impact of the EDM constraints enough to open up some of the parameter space for stop-sourced EWB.  However, although the EDM constraints are weakened, the baryon asymmetry is also reduced and once again the $^{199}$Hg constraints rule out all available parameter space.  Finally, increasing both $\left|A_t\right|$ and $\mu$ to 1000 GeV (see Fig.~\ref{fig:stop_sources2}) yields larger regions excluded by direct stop searches (due to the large mixing) and weaker Higgs constraints, allowing one to push $m_{h_0}$ up to 120 GeV in parts of the parameter space.  The EDM constraints in this case are more stringent, again ruling out all available parameter space for stop-sourced EWB.  We note that in addition to the scenarios shown in Figs.~\ref{fig:stop_sources1}-\ref{fig:stop_sources2} we could have also chosen a small value of $\mu$ and large $\left|A_t\right|$, however in this case the EDM constraints are again very stringent and all the stop-sourced EWB parameter space is excluded; we omit the corresponding figures for brevity (note that in this case one can obtain larger Higgs masses).  We have also verified that decreasing $\sin\phi_t$ does not open up any more parameter space for stop sources.  

Additionally, varying $\tan \beta$ and/or $m_A$ does not affect our results.  Since the stop-sourced baryon asymmetry and EDM predictions for light stop contributions scale with $y_t\sim1/\sin\beta$, they are both rather insensitive to changes in $\tan\beta$ (we return to the large-$\tan \beta$ regime in the following section). 
Increasing $m_A$ suppresses both the overall baryon asymmetry and the expected EDMs, but the BAU varies as $1/m_A^2$ whereas the EDMs only vary as $1/m_A$, so a heavier CP-odd Higgs provides stronger exclusions.  Smaller values of $m_A$ can enhance the BAU up to about a factor of $4$ (for $m_A=100$ GeV), but we have checked that this does not overcome the strong EDM exclusions in any of the cases considered. 

As discussed above, the computation of $d_{Hg}$ involves significant theoretical uncertainty, which could impact the strength of the above conclusions. However, even if the theoretical prediction for $d_{Hg}$ were in fact an order of magnitude smaller than the values used in Figs.~\ref{fig:stop_sources1}, \ref{fig:stop_sources2}, it would still be just as constraining as the electron-EDM, which by itself rules out virtually all of the parameter space with $|Y_B| \geq Y_{Obs}$. We expect the vev-insertion approximation to \emph{over}-estimate the produced baryon asymmetry, and so it is unlikely that even this large correction would in fact open up any additional space for stop sources in a more careful treatment beyond the approximations used here.  Similar considerations hold for the sbottom sources in Sec.~\ref{sec:largetanbeta} as well.

Summarizing, in considering the various scenarios depicted in Figs.~\ref{fig:stop_sources1}-\ref{fig:stop_sources2}, \emph{we find no viable parameter space for MSSM stop-driven resonant EWB consistent with Higgs mass, stop mass, and EDM constraints.}  Even conservatively estimating the result of the various uncertainties of the calculations as increasing $Y_B$ by an order of magnitude does not open up any viable parameter space for stop-sourced EWB.  The large experimentally excluded regions are primarily a result of the stringent EDM constraints, and particularly that of $^{199}$Hg.  We have also verified that even e.g. a factor of ten decrease in the Mercury EDM limits does not open up any additional parameter space for the stops.  It is difficult to see how one might circumvent these constraints to produce the correct baryon asymmetry through a scenario relying primarily on stop sources.

\section{The Large $\tan\beta$ Regime: Sbottom and Stau Sources}\label{sec:largetanbeta}

We now turn our attention to the other third-generation scalars as a possible source for the observed baryon asymmetry.  Since the CP-violating sources arising from triscalar interactions for the sbottoms and staus are proportional to their Yukawa couplings, for these sources to contribute significantly to the BAU, one must consider larger values of $\tan \beta$.  For large $\tan \beta$, the sbottom and stau Yukawa couplings, $y_b$, $y_{\tau}$, are enhanced as \begin{equation} \label{eq:yukawas}  y_b=\frac{m_b}{v \cos\beta}, \qquad y_{\tau}=\frac{m_{\tau}}{v \cos \beta}, \end{equation} where $v\approx 175$ GeV is the Higgs vev at $T=0$.  In what follows, we take $\tan \beta =40$ so that the strength of the sbottom and stau CP-violating sources are effectively comparable to that of the stops.

\subsection{Sbottoms}
To compute the CP-violating source for resonant sbottom scattering off of the EWPT bubble wall, we make use of the Lagrangian in Eq.~(\ref{eq:lagrangian}).  The sbottom interaction Lagrangian differs from that of the stops by the replacements $\left\{\tilde{t}, A_t, y_t, v_u, v_d\right\} \rightarrow \left\{ \tilde{b}, A_b, y_b, v_d, v_u\right\}$ and the relevant CP-violating phase is $\phi_b=\mathrm{Arg}(\mu A_b b^\ast)$.  One can use these replacements in the non-equilibrium field theory derivation for $S_{\tilde{t}}^{CPV}$ in Ref.~\cite{Lee:2004we} to obtain $S_{\tilde{b}}^{CPV}$.  The resulting CP-violating sbottom source is given by \begin{equation} \label{eq:sbottom_source} \begin{aligned} S_{\tilde{b}}^{CPV}(x)=&-\frac{N_Cy_b^2}{2\pi^2}\operatorname{Im}(\mu A_b)v^2(x)\dot{\beta}(x) \\ &\times \int_0^{\infty} \frac{dk k^2}{\omega_R \omega_L}  \operatorname{Im}\left[\frac{n_B(\mathcal{E}^*_R)-n_B(\mathcal{E}_L)}{(\mathcal{E}_L-\mathcal{E}_R^*)^2}+\frac{n_B(\mathcal{E}_R)+n_B(\mathcal{E}_L)}{(\mathcal{E}_L+\mathcal{E}_R)^2} \right] \end{aligned}\end{equation} where the $\mathcal{E}_{L,R}$, $\omega_{L,R}$ terms are as in Eq.~(\ref{eq:definitions}) with $M_{\tilde{t}}\rightarrow M_{\tilde{b}}$ and $\Gamma_i$ corresponding to the thermal widths of the LH- and RH-sbottoms.  Notice that the coupling of the sbottom to the down-type Higgs vev manifests itself as an overall relative sign between $S_{\tilde{b}}^{CPV}$ and $S_{\tilde{t}}^{CPV}$.  The source enters into the same set of QBEs, Eqs.~(\ref{QBE1})-(\ref{eq:QBE_last}), and since there is now a source for $\tilde{b}_R$, one must include the density $Q_1$ in the network of equations.  Since we take the sleptons to be heavy and the SM leptonic Yukawa interaction rates are small compared to the corresponding rates for the quarks, we neglect $\Gamma_{y\tau}$ in our calculation of the sbottom-sourced baryon asymmetry.  As a result, the equations for the densities $L$ and $R$ decouple from the full set of QBEs; the relevant set of Boltzmann equations to solve is then given by Eqs.~(\ref{QBE1}), (\ref{QBE2}), (\ref{QBE3}) and  (\ref{QBE4}) with the replacement $\Gamma_{y\tau}\rightarrow 0$.  In terms of the relevant charge densities, the left-handed fermionic charge density in Eq.~(\ref{eq:nL}) simplifies to \begin{equation} \label{eq:nL_2}n_L=\frac{k_q}{k_Q}Q+2Q_1. \end{equation} We note that since $\tilde{t}_R$ is heavy, the right-handed stops and tops are no longer in superequilibrium.  This manifests itself in the contributions to the Yukawa and relaxation rates involving the RH stops vanishing, while the density $T$ in Eq.~(\ref{QBE1}) corresponds entirely to a SM charge density.  In the parameter space we consider the (s)bottoms are in superequillibrium everywhere except the kinematically disallowed region for the RH rate, and we proceed analogously to the (s)top case by assuming $\mu_{\tilde{b}_{L.R}}=\mu_{b_{L,R}}$ superequilibrium in all the parameter space when computing the baryon asymmetry.

%%%%%%%%%%%%%%%%%%%%%%%%%%%%%%%%%%%%%%%%%%%%%%%%%%%
\begin{figure*}[!t]
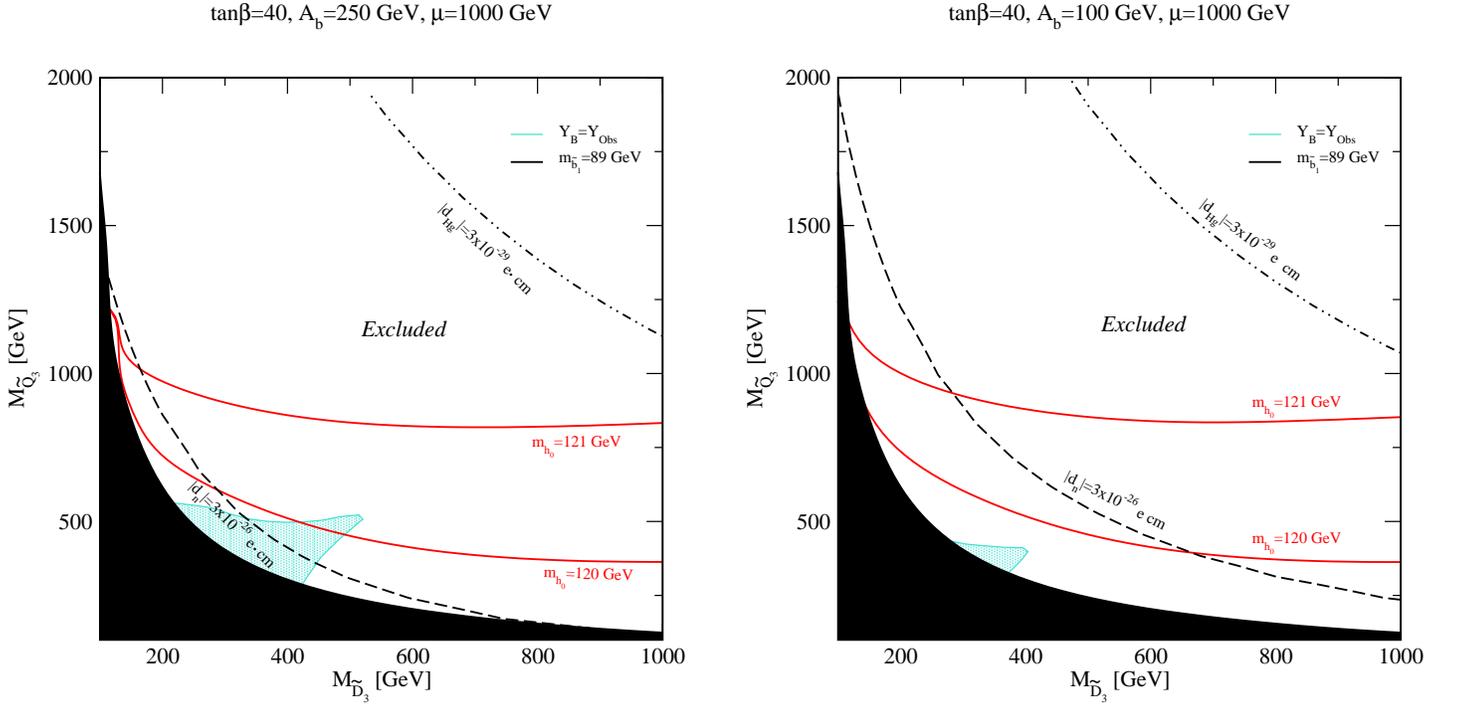

\mbox{\hspace*{-1.2cm}\includegraphics[width=0.55\textwidth,clip]{definitive_plots/sbottoms_tanb_Ab_mu_40_250_1000.eps}\qquad\includegraphics[width=0.55\textwidth,clip]{definitive_plots/sbottoms_tanb_Ab_mu_40_100_1000.eps}}
\caption{\label{fig:sbottom_sources1}\it\small Regions of the sbottom soft supersymmetry breaking mass parameter space consistent with the observed value of the baryon asymmetry resulting from sbottom sources for $\mu=1000$ GeV, $\left|A_b\right|=250$ GeV (Left) and $\left|A_b\right|=100$ GeV (Right).  Regions shaded blue correspond to $Y_B\geq Y_{Obs}$ for maximal CP-violating phase.  The curve corresponding to an overestimate of $Y_B$ by a factor of 10 falls within the black shaded regions, which are excluded by sbottom mass direct searches.  Red lines are iso-contours of the SM-like Higgs mass $m_{h_0}$; the LEP bound is satisfied in all of the parameter space shown.  The current constraint on the neutron and Mercury EDMs are represented by the black dashed and dashed-double dot lines, respectively, with regions to the left of each line ruled out by null results.  For the $\left|A_b\right|=250$ GeV case, the current $d_e$ bound falls beneath the shaded region while the current constraint on the Mercury EDM rules out all of the parameter space shown.  Future EDM measurements of $d_e$, $d_n$, and $d_{Hg}$ are expected to definitively probe well beyond the shown parameter space.  For the $\left|A_b\right|=100$ GeV case, the electron EDM current bound falls beneath the black shaded region, while future EDM measurements of $d_e$, $d_n$, and $d_{Hg}$ will again probe all of the parameter space shown.}
\end{figure*}
%%%%%%%%%%%%%%%%%%%%%%%%%%%%%%%%%%%%%%%%%%%%%%%%%%%

%%%%%%%%%%%%%%%%%%%%%%%%%%%%%%%%%%%%%%%%%%%%%%%%%%%
\begin{figure*}[!t]
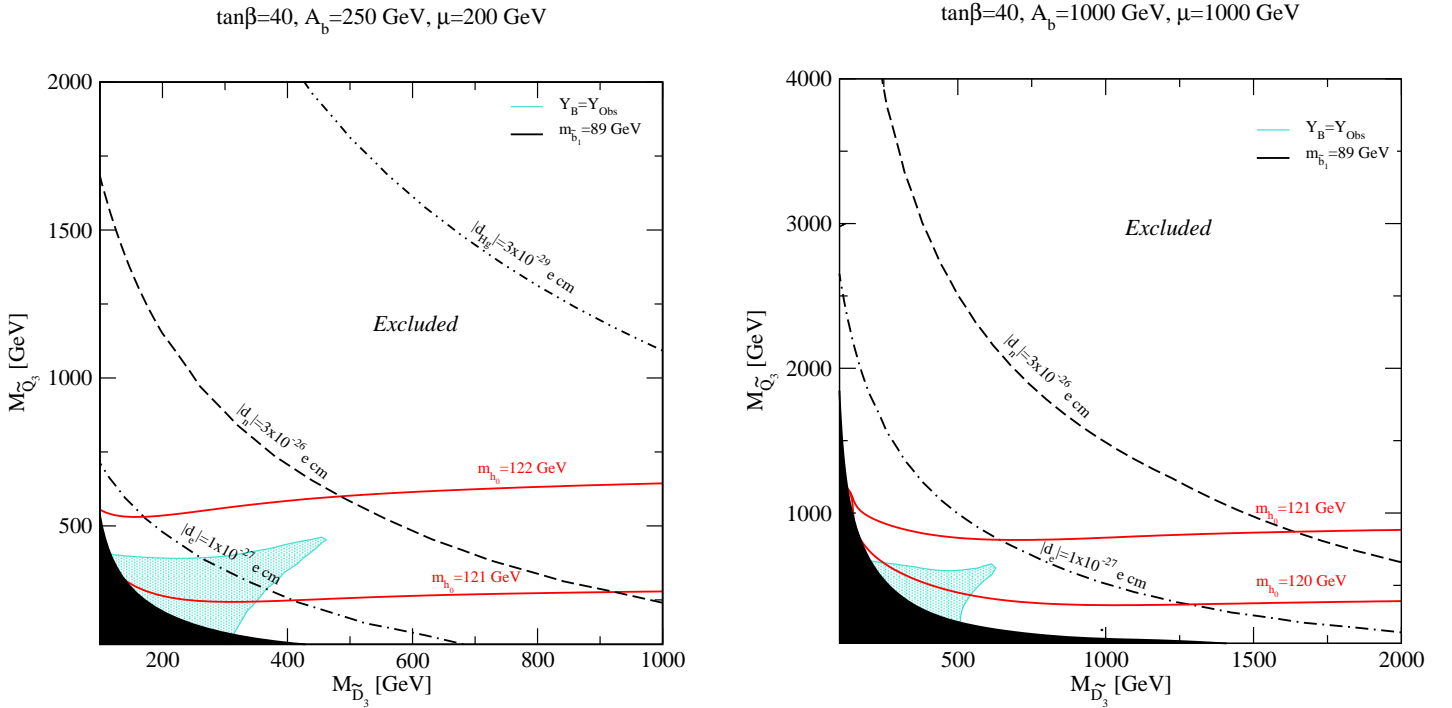

\mbox{\hspace*{-1.2cm}\includegraphics[width=0.55\textwidth,clip]{definitive_plots/sbottoms_tanb_Ab_mu_40_250_200.eps}\qquad\includegraphics[width=0.55\textwidth,clip]{definitive_plots/sbottoms_tanb_Ab_mu_40_1000_1000.eps}}
\caption{\label{fig:sbottom_sources2}\it\small Same as Fig.~\ref{fig:sbottom_sources1} but for $\left|A_b\right|=250$ GeV, $\mu=200$ GeV (Left) and $\left|A_b\right|=1000$ GeV, $\mu=1000$ GeV (Right).  For both the $\left|A_b\right|=250$ and $\left|A_b\right|=1000$ GeV cases, the future reach of electron, neutron, and Mercury EDM measurements is expected to probe the entire parameter space shown.  For the $\left|A_b\right|=1000$ GeV case, the current Mercury EDM constraint  already rules out all of the parameter space shown.}
\end{figure*}
%%%%%%%%%%%%%%%%%%%%%%%%%%%%%%%%%%%%%%%%%%%%%%%%%%%

We calculate $Y_B$ following the spectrum outlined in Sec.~\ref{sec:params}, only now with $100$ GeV$\le M_{\tilde{D}_3}\le 2000$, $M_{\tilde{U}_3}=10$ TeV, $A_t=0$, and varying $\left|A_{b}\right|= 100$, 250, 1000 GeV.  As for the stops, we assume that a strongly first order phase transition is generated from some mechanism other than the light stop scenario.  The resulting sbottom-sourced BAU is plotted in Figs.~\ref{fig:sbottom_sources1}-\ref{fig:sbottom_sources2}, where regions compatible with the observed asymmetry are shaded blue (we find no sign change in $Y_B$ for the sbottoms with our choices of parameters).  The resonance in the CP-violating source is again apparent.  

In Figs.~\ref{fig:sbottom_sources1}-\ref{fig:sbottom_sources2} we also show the lower bound on the sbottom mass from direct searches, $m_{\tilde{b}_1}\gtrsim 89$ GeV \cite{PDG} and contours of constant SM-like Higgs mass.  The LEP bound on $m_{h_0}$ is satisfied in all regions of parameter space considered.  Since the mass of the RH stop is heavy, $m_{h_0}$ receives larger contributions from stop loops compared to the stop-source case and one can easily push the Higgs mass up to $m_{h_0}\approx 120$ GeV, however larger values are more difficult to obtain with our choices of parameters (as with the stops, these should not be taken as strict exclusions).  The EDM constraints for the sbottom sources are similar to those for the stop sources, but they receive a $\tan \beta$ enhancement, and thus the constraints tend to be more stringent. 
  
The behavior of the produced baryon asymmetry and the various constraints in Figs.~\ref{fig:sbottom_sources1}-\ref{fig:sbottom_sources2} is qualitatively similar to that for the stop-source case: increasing $\left|A_b\right|$ or $\mu$ leads to larger regions compatible with the observed baryon asymmetry but strengthens the various EDM constraints.  We note that since the sbottoms have down-type couplings to the Higgs, the roles of $A_b$ and $\mu$ in the mass matrix for the $T=0$ sbottoms are reversed relative to the roles of $A_t$ and $\mu$ for the stops, and as a result, the exclusions from direct searches are primarily sensitive to $\mu$ for the large value of $\tan \beta$ chosen here.  In addition to the cases shown in Figs.~\ref{fig:sbottom_sources1}-\ref{fig:sbottom_sources2}, we have verified that scenarios for sbottom-sourced EWB with large $\left|A_b\right|$ and small $\mu$ are also solidly ruled out by the current $^{199}$Hg EDM constraint.  We have also checked that decreasing the strength of the CP-violating phase opens up no additional parameter space for the sbottom sources (it potentially could have, as the slope of EDM constraints on the shown parameter space is different from that of BAU isolevel curves).  Consequently, \emph{taking sbottom mass, Higgs mass, and EDM constraints into account, we find no regions of the sbottom mass parameter space consistent with the observed value of $Y_B$: as for stop sources, current EDM constraints imply that sbottom sources alone cannot explain the BAU in the context of SUSY EWB}.  

\subsection{Staus}

Finally, we consider the case where the observed baryon asymmetry may have arisen primarily from CP-violation in the stau sector of the MSSM.  For large values of $\tan\beta$, $y_{\tau}$ can become enhanced as per  Eq.~(\ref{eq:yukawas}).  From the Lagrangian, Eq.~(\ref{eq:lagrangian}), and following Ref.~\cite{Lee:2004we}, we can proceed in parallel to the  calculation of Eq.~(\ref{eq:sbottom_source}) for the CP-violating stau source $S_{\tilde{\tau}}^{CPV}$ with the replacements $\{\tilde{b}, A_b, y_b\} \rightarrow \{ \tilde{\tau}, A_{\tau}, y_{\tau}\}$, yielding 
\begin{equation}  
 \label{eq:stau_source} 
 \begin{aligned}
 S_{\tilde{\tau}}^{CPV}(x)=&-\frac{y_{\tau}^2}{2\pi^2}\operatorname{Im}(\mu A_{\tau})v^2(x)\dot{\beta}(x)\\& \times \int_0^{\infty} \frac{dk k^2}{\omega_R \omega_L}  \operatorname{Im}\left[\frac{n_B(\mathcal{E}^*_R)-n_B(\mathcal{E}_L)}{(\mathcal{E}_L-\mathcal{E}_R^*)^2}+\frac{n_B(\mathcal{E}_R)+n_B(\mathcal{E}_L)}{(\mathcal{E}_L+\mathcal{E}_R)^2} \right] 
 \end{aligned}
\end{equation} 
and with the appropriate replacements in the definitions of Eq.~(\ref{eq:definitions}) for the LH and RH staus.  The relevant CP-violating phase is now $\phi_\tau=\mathrm{Arg}(\mu A_\tau b^\ast)$.  The source Eq.~(\ref{eq:stau_source}) enters the full set of QBEs, since for large $\tan \beta$ all third-generation Yukawa rates should be taken into account.  The left-handed fermionic charge density is given by \begin{equation}\label{nL_3} n_L=Q+2Q_1+\frac{k_l}{k_L} L \end{equation} where $k_l$ is the fermionic contribution to $k_L$.  Note that unlike for quarks, only the third generation LH density $L$ contributes to $n_L$ since there is no generational mixing for leptons and we have neglected the first- and second- generation leptonic Yukawa couplings.  We have verified that the staus and taus are in superequilibrium everywhere except in kinematically disallowed regions, so we proceeded as before, assuming $\mu_{\tilde{\tau}_{L,R}}=\mu_{\tau_{L,R}}$ in computing $Y_B$.

\begin{figure*}[!t]
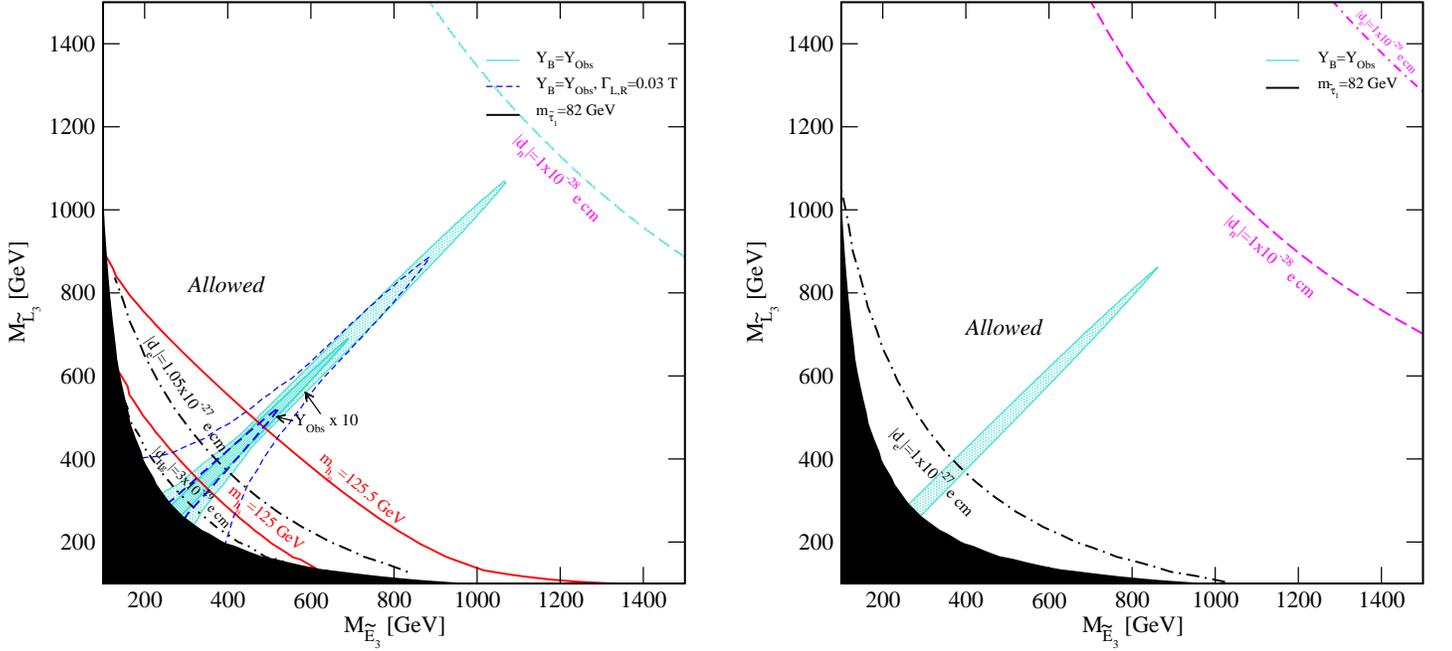

\mbox{\hspace*{-1.2cm}\includegraphics[width=0.55\textwidth,clip]{definitive_plots/comparison_stau_tan_Atau_mu_40_250_1000.eps}\qquad\includegraphics[width=0.55\textwidth,clip]{definitive_plots/stau_tan_Atau_mu_40_100_1000.eps}}
\caption{\label{fig:stau_sources1}\it\small As in Figs.~\ref{fig:stop_sources1}-\ref{fig:stop_sources2}, but for stau sources.  For the $\left|A_{\tau}\right|=250$ GeV case, the dashed blue lines correspond to constant-$Y_B$ curves computed for a factor of ten larger thermal stau widths.  In this case, the expected reach of future $d_e$ measurements will probe all of the parameter space shown.  For the $\left|A_{\tau}\right|=100$ GeV case, neutron and Mercury EDM bounds fall beneath the black shaded region.  In both cases the expected reach of future $d_{Hg}$ measurements is nearly degenerate with the current bound from measurements of $d_e$ and is not shown.}
\end{figure*}

\begin{figure*}[!t]
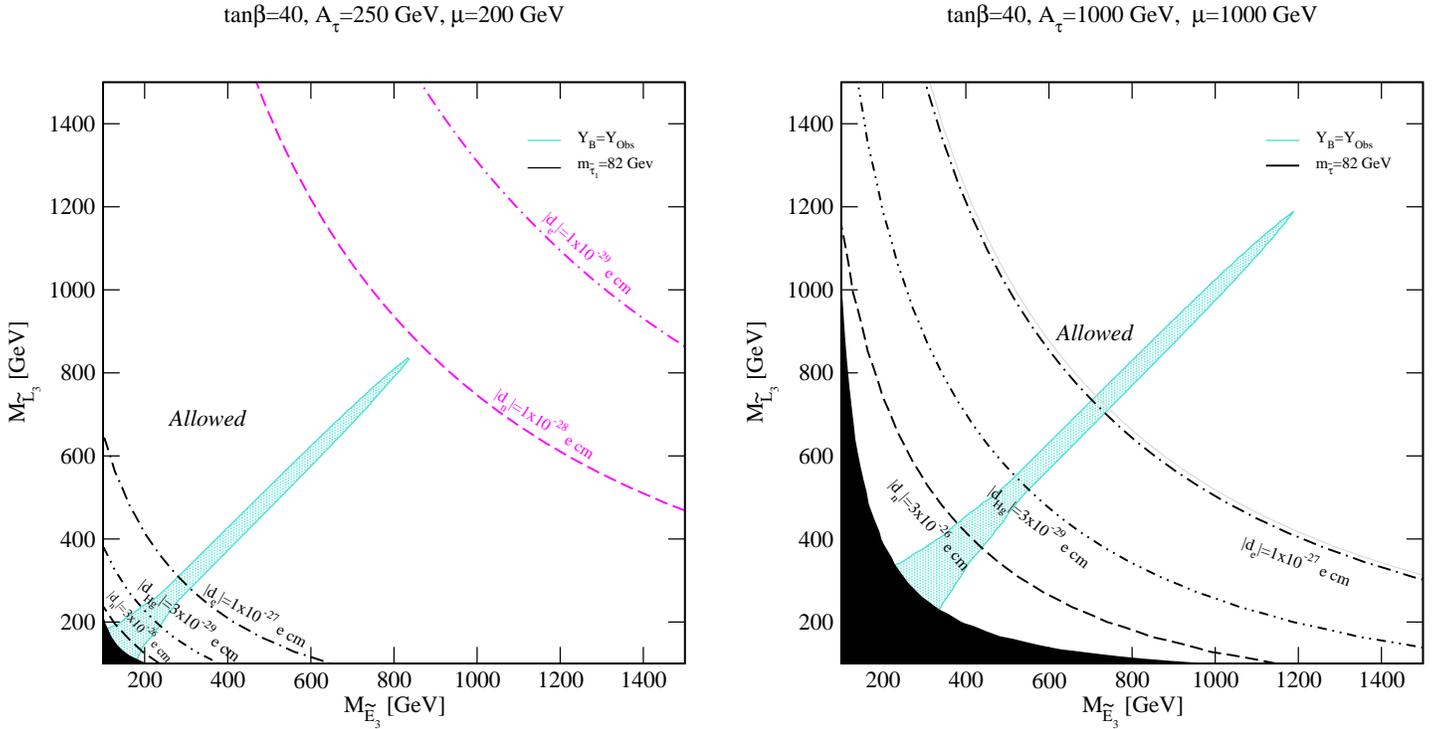

\mbox{\hspace*{-1.2cm}\includegraphics[width=0.55\textwidth,clip]{definitive_plots/stau_tan_Atau_mu_40_250_200.eps}\qquad\includegraphics[width=0.55\textwidth,clip]{definitive_plots/stau_tan_Atau_mu_40_1000_1000.eps}}
\caption{\label{fig:stau_sources2}\it\small  As in Fig.~\ref{fig:stau_sources1} for different values of $A_{\tau}$, $\mu$.  For the $\left|A_{\tau}\right|=1000$ GeV case,  future electron and neutron EDM experiments will probe all parameter space shown.  In both cases the expected reach of future $d_{Hg}$ measurements is nearly degenerate with the current bound from measurements of $d_e$ and is not shown here}
\end{figure*}

For the spectrum we again proceed in parallel to the analysis outlined in Sec.~\ref{sec:params} with the appropriate replacements for the staus, again assuming a strongly first-order phase transition, either from the light stop scenario or some other mechanism (in calculating the BAU and constraints we assume a heavy RH stop).  The resulting slepton-sourced baryon asymmetry is shown in Figs.~\ref{fig:stau_sources1}-\ref{fig:stau_sources2} for various values of $\left|A_{\tau}\right|$, $\mu$ and maximal CP-violating phase $\phi_{\tau}$; regions of the stau mass parameter space compatible with successful EWB are shaded blue.  The resulting baryon asymmetry is strongly peaked near the resonance.  This is because the thermal widths of the staus in the plasma, which enter into $\mathcal{E}_{L,R}$ in the denominator of $S_{\tilde{\tau}}^{CPV}$, are much smaller than those for the squarks, $\Gamma_{Q,T}\simeq0.5 T$, $\Gamma_{L,R}\simeq 0.003 T$ \cite{Chung:2009qs}.  As a result, successful stau-sourced EWB requires nearly degenerate $\tilde{\tau}_L$, $\tilde{\tau}_R$; from Figs.~\ref{fig:stau_sources1}-\ref{fig:stau_sources2} we find $\left|M_{\tilde{L}_3}-M_{\tilde{E}_3}\right|\lesssim 100$ GeV to produce the observed value of $Y_B$ for $\sin \phi_{\tau}=1$.  

%%%%%%%%%%%%%%%%%%%%%%%%%%%%%%%%%%%%%%%%%%%%%%%%%%%
%%%%%%%%%%%%%%%%%%%%%%%%%%%%%%%%%%%%%%%%%%%%%%%%%%%

%%%%%%%%%%%%%%%%%%%%%%%%%%%%%%%%%%%%%%%%%%%%%%%%%%%
%%%%%%%%%%%%%%%%%%%%%%%%%%%%%%%%%%%%%%%%%%%%%%%%%%%

The results shown in Figs.~\ref{fig:stau_sources1}-\ref{fig:stau_sources2} demonstrate that the resonance supplied by the small thermal widths of the staus present in the denominator of Eq.~\ref{eq:stau_source} can overcome the suppression effect of the resonant relaxation rate $\Gamma_{m\tau}$.  This can be understood by noting that the overall baryon asymmetry scales with \cite{Lee:2004we} $\sim S^{CPV}_{\tilde{\tau}}/\sqrt{\Gamma_{m\tau}}$ so although both the source and relaxation rates are resonantly enhanced by the small widths, the asymmetry will tend to increase with decreasing widths.  Also, the strong resonance in the denominator of $S^{CPV}_{\tilde{\tau}}$ can overcome the Boltzmann suppression in the numerator for stau soft masses up to $\sim 1$ TeV in most cases.  Physically, this corresponds to the very efficient production of chiral charge by a relatively small abundance of staus in the plasma.  These results hinge on the small values of $\Gamma_{L,R}=0.003$, which we take from Ref.~\cite{Chung:2009qs} and which were computed for $\tan\beta=15$.  One might expect the widths to be enhanced for the larger values of $\tan\beta$ we consider here, since e.g. the otherwise negligible Yukawa decay $\tilde{\tau}\rightarrow \tilde{H} \tau$ can become important in this regime, yielding an enhancement of $\Gamma_{L,R}$ from this extra decay channel of a factor of order 2 at most.  Also, $\Gamma_{L,R}$ are not necessarily equal, due to the differing hypercharges in the decays $\tilde{\tau}\rightarrow \tau \tilde{B}$, as well as the $SU(2)$ decay to $\tau \tilde{W}$ which can be open for our choices of gaugino masses.  To demonstrate the sensitivity of our results to the precise values of the thermal widths, we include on the LHS of Fig.~\ref{fig:stau_sources1} the curves of $Y_{\rm Obs}$, $10\times Y_{\rm Obs}$ calculated for an order-of-magnitude larger widths, $\Gamma_{L,R}=0.03 T$, which we expect to over-estimate the uncertainty in $\Gamma_{L,R}$ associated with these considerations.  Even this factor of $10$ increase in $\Gamma_{L,R}$ admits a significant amount of parameter space compatible with stau-sourced EWB. We thus expect that our overall conclusions are rather insensitive to the details entering into a more precise determination of the stau thermal widths, however we encourage the Reader to keep the above caveats in mind when interpreting our results. 

In Figs.~\ref{fig:stau_sources1}-\ref{fig:stau_sources2} we also plot constraints from direct searches for staus, $m_{\tilde{\tau}_1}\gtrsim 82$ GeV \cite{PDG}, which display the down-type dependence on $\left|A_{\tau}\right|$ and $\mu$ similar to that of the sbottoms.  We also show iso-level contours of constant Higgs mass for the $\left|A_{\tau}\right|=250$ GeV, $\mu=1000$ GeV case.  We omit these curves for the other plots since the exact values of the SM-like Higgs mass in each case are sensitive to the details of e.g. squark and gluino masses which do not impact the determination of $Y_B$ in slepton-sourced EWB.  Two significant features emerge: 
\begin{itemize}
\item[(1)]From Fig.~\ref{fig:stau_sources1} we see that one can achieve a  Higgs mass in this scenario consistent with the hints from ATLAS and CMS, $m_{h_0}\sim 125$ GeV.  In contrast to our analysis of the stop and sbottom sources wherein we found no viable regions of parameter space for  $m_{h_0}\gtrsim 120$ GeV, we are able to easily obtain a heavier Higgs mass for the stau source case since we are free  to consider heavy squarks which contribute large loop corrections to $m_{h_0}$.  Also note that the excess events observed in the $H\to\gamma\gamma$ channel with respect to general MSSM expectations \cite{higgsclaims} could favor a scenario with light staus \cite{carlos}.
\item[(2)] Alternatively, one may also obtain the correct baryon asymmetry from stau sources along with a light right-handed stop; since there is no CP-violating phase in $A_t$, the EDM constraints will not be affected, however the large loop contributions to $m_{h_0}$ will be lost. 
\end{itemize}

We also consider EDM constraints on the stau-source scenario in Figs.~\ref{fig:stau_sources1}-\ref{fig:stau_sources2}.  In order to generate chromo-EDMs, one needs  a CP-violating phase that couples to (s)quarks. When the only phase is $\phi_\tau$, the chromo-EDMs disappear. Consequently, both the neutron and Mercury EDM constraints are much weaker in this scenario, while the electron EDM is the relevant one. The electron EDM in the case of CP violation in the stau sector entirely stems from a single Barr--Zee graph with a stau loop.  From Figs.~\ref{fig:stau_sources1}-\ref{fig:stau_sources2}, we see that the lack of chromo-EDMs opens up large sections of parameter space, allowing for viable baryogenesis while satisfying the experimental constraints.  Future EDM experiments are expected to probe all of the parameter space available for stau-mediated EWB.  Note that since the primary constraint on stau sources is the electron EDM, the available parameter space is rather insensitive to the theoretical uncertainty in the calculation of $d_{Hg}$: an order of magnitude \emph{under}-estimate of the mercury EDM would make its constraints comparable to that of the electron EDM.  This picture holds even for smaller values of the CP-violating phase; we find that one can produce the correct BAU and still satisfy the various constraints for $\sin \phi_{\tau}\gtrsim 10^{-2}$ in most cases considered.

Summarizing the results of Figs.~\ref{fig:stau_sources1}-\ref{fig:stau_sources2}, \emph{we conclude that it is possible to produce the observed baryon asymmetry with CP violation in the stau sector only, for nearly degenerate staus and $300$ GeV $\lesssim M_{\tilde{L}_3, \tilde{E}_3}\lesssim 1.2$ TeV, depending on the magnitudes of the stau triscalar coupling  and $\mu$, while satisfying EDM and direct search constraints.  This scenario can also naturally accommodate an SM-like Higgs mass $m_{h_0}\sim 125$ GeV for heavy squarks, or possibly a strongly first order phase transition via the light stop scenario for light $m_{\tilde{t}_1}$.}  Should future searches for electron, neutron, and Mercury EDMs yield null results, all scalar sources in the MSSM will be ruled out for resonant EWB.

\section{Discussion and Conclusions}\label{sec:disc}

The exploration of the electroweak scale is closing in on supersymmetric models in which the baryon asymmetry is produced at the electroweak phase transition. In this study, we have focused on the possibility that the CP violation necessary for successful EWB is found in the third generation sfermion sector, a scenario considered only for stops in the past. Here we have studied, for the first time, CP violation in the third generation {\em down-type} sfermion sector, i.e. sbottoms and staus, as a source relevant for baryogenesis. Also, we have quantitatively addressed the question of whether or not sfermionic CP violating sources are compatible with current constraints on the size of the electron, neutron and atomic electric dipole moments.

The main findings of this study are
\begin{itemize}
\item[(1)]Neither the stop nor the sbottom sector are viable options to account for the bulk of the observed baryon asymmetry of the universe: two-loop Barr-Zee diagrams contribute to the chromo-EDM of the down quark to a level that is ruled out by current constraints of the Mercury EDM across the entirety of the parameter space where stop or sbottom sources could source a large enough amount of baryon asymmetry. Moreover, stop- and/or sbottom-mediated EWBG is disfavored by indications of $m_{h_0}\approx 125$ GeV, though present Higgs search constraints on the CP-violating sources are not nearly as decisive as those arising from EDMs.
\item[(2)] The stau sector (where no chromo-EDMs are produced) has milder constraints from EDMs, and hence can be responsible for producing the net left-handed chiral charge density needed to produce, via weak sphaleron transitions, the observed baryon asymmetry in the universe. It is also possible in this case to achieve $m_{h_0}\approx 125$ GeV or a light RH stop as needed for a strong first order phase transition, but not both. Due to the relatively small stau thermal widths, however, this scenario of ``slepton-mediated'' electroweak baryogenesis requires almost degenerate staus, with masses between 300 GeV and 1.2 TeV, depending on the size of the stau triscalar coupling and $\mu$. This scenario also requires large values of $\tan\beta$.
\end{itemize}

While, from the standpoint of requiring a strongly first order phase transition, electroweak baryogenesis in the MSSM is being conclusively tested by ongoing searches for a light stop and the Higgs \cite{Curtin:2012aa}, the present results provide important, complementary information on the nature of the new sources of CP violation, the second key ingredient for successful supersymmetric electroweak baryogenesis. Future results from the LHC and from the next generation of EDM searches are therefore expected to yield an increasingly sharper, if not definitively clear, picture of whether or not the electroweak scale is related to the generation of the observed baryon asymmetry.

\begin{acknowledgments}
\noindent  SP and JK are partly supported by an Outstanding Junior Investigator Award from the US Department of Energy and by Contract DE-FG02-04ER41268, and by NSF Grant PHY-0757911. CLW is supported by an NSF graduate fellowship. MJRM is supported by US Department of Energy and by Contract DE-FG02-08ER41531 and the Wisconsin Alumni Research Foundation.

\end{acknowledgments}

\end{document}